\documentclass[sigconf]{acmart}
\renewcommand\footnotetextcopyrightpermission[1]{} 
\usepackage[utf8]{inputenc}
\usepackage{algpseudocode,algorithm,algorithmicx}
\usepackage[nolist,nohyperlinks]{acronym}
\usepackage{booktabs}
\usepackage[binary-units = true]{siunitx}
\usepackage{csvsimple}
\usepackage[capitalise]{cleveref}

\newcommand*{\symDefine}[2]{\newcommand{{#1}}{{#2}}}
\symDefine{\symPrecision}{p}
\symDefine{\symRegRange}{q}
\symDefine{\symRegVal}{k}
\symDefine{\symRegValVariate}{K}
\symDefine{\symNumReg}{m}
\symDefine{\symDataItem}{D}
\symDefine{\symDistanceMeasure}{D}
\symDefine{\symBitRepA}{a}
\symDefine{\symBitRepB}{b}
\symDefine{\symNumBitsMinwiseHashing}{b}
\symDefine{\symImaginary}{\mathrm{i}}
\symDefine{\symIndexI}{i}
\symDefine{\symIndexJ}{j}
\symDefine{\symIndexL}{l}
\symDefine{\symCount}{c}
\symDefine{\symCountVariate}{C}
\symDefine{\symAlpha}{\alpha}
\symDefine{\symBeta}{\beta}
\symDefine{\symA}{a}
\symDefine{\symB}{b}
\symDefine{\symS}{s}
\symDefine{\symX}{x}
\symDefine{\symY}{y}
\symDefine{\symZ}{z}
\symDefine{\symPhi}{\varphi}
\symDefine{\symCardinality}{n}
\symDefine{\symCardinalityEstimate}{\hat{\symCardinality}}
\symDefine{\symCardinalityRawEstimate}{\symCardinalityEstimate_\text{raw}}
\symDefine{\symError}{\varepsilon}
\symDefine{\symStopDelta}{\delta}
\symDefine{\symStopEpsilon}{\varepsilon}
\symDefine{\symBigO}{\mathcal{O}}
\symDefine{\symExpectation}{\mathbb{E}}
\symDefine{\symProbability}{P}
\symDefine{\symProbabilityMass}{\rho}
\symDefine{\symRegProbability}{\gamma}
\symDefine{\symLikelihood}{\mathcal{L}}
\symDefine{\symPoissonRate}{\lambda}
\symDefine{\symPoissonRateEstimate}{\hat{\symPoissonRate}}
\symDefine{\symFunc}{f}
\symDefine{\symFuncPrime}{g}
\symDefine{\symSetA}{A}
\symDefine{\symSetB}{B}
\symDefine{\symSetS}{S}
\symDefine{\symSetX}{X}
\symDefine{\symSetU}{U}
\symDefine{\symSetASuffix}{a}
\symDefine{\symSetBSuffix}{b}
\symDefine{\symSetXSuffix}{x}
\symDefine{\symSetUSuffix}{u}
\symDefine{\symPowerSeriesFunc}{\xi}
\symDefine{\symSmallCorrectionFunc}{\sigma}
\symDefine{\symLargeCorrectionFunc}{\tau}
\symDefine{\symEpsPowerSeriesFunc}{\eta}
\symDefine{\symBiasCorrectionFunc}{w_{\text{corr}}}

\newcommand{\numformat}[1]{{\num[scientific-notation = true,round-mode = places,round-precision = 2, output-exponent-marker = \ensuremath{\mathrm{E}}]{#1}}}

\newcommand{\numformattwo}[1]{{\num[scientific-notation = fixed,round-mode = places,round-precision = 2,round-integer-to-decimal=true]{#1}}}

\setcopyright{none}

\begin{document}
\sloppy
\allowdisplaybreaks
\title{New Cardinality Estimation Methods for HyperLogLog Sketches}

\author{Otmar Ertl}
\affiliation{
  \city{Linz} 
  \country{Austria} 
}
\email{otmar.ertl@gmail.com}

\begin{abstract}
This work presents new cardinality estimation methods for data sets recorded by HyperLogLog sketches. A simple derivation of the original estimator was found, that also gives insight how to correct its deficiencies. The result is an improved estimator that is unbiased over the full cardinality range, is easy computable, and does not rely on empirically determined data as previous approaches. 
Based on the maximum likelihood principle a second unbiased estimation method is presented which can also be extended to estimate cardinalities of union, intersection, or relative complements of two sets that are both represented as HyperLogLog sketches. Experimental results show that this approach is more precise than the conventional technique using the inclusion-exclusion principle.
\end{abstract}

\maketitle

\begin{acronym}
\acro{HLL}{HyperLogLog}
\acro{ML}{maximum likelihood}
\acro{BFGS}{Broyden-Fletcher-Goldfarb-Shanno}
\acro{RMSE}{root-mean-square error}
\end{acronym}

\section{Introduction}
Counting the number of distinct elements in a data stream or large datasets is a common problem in big data processing. 
In principle, finding the number of distinct elements $\symCardinality$ with a maximum relative error $\symError$  in a data stream requires $\symBigO(\symCardinality)$ space \cite{Alon1999}. However, probabilistic algorithms that achieve the requested precision only with high probability are able to drastically reduce space requirements. Many different probabilistic algorithms have been developed over the past two decades \cite{Metwally2008,Ting2014} until a theoretically optimal algorithm was finally found \cite{Kane2010}. Although this algorithm achieves the optimal space complexity of $\symBigO(\symError^{-2}+\log \symCardinality)$ \cite{Alon1999, Indyk2003}, it is not very efficient in practice \cite{Ting2014}.

More practicable and already widely used in many applications is the \ac{HLL} algorithm \cite{Flajolet2007} with a near-optimal space complexity $\symBigO(\symError^{-2} \log\log\symCardinality +\log \symCardinality)$. A big advantage of the \ac{HLL} algorithm is that corresponding sketches can be easily merged, which is a requirement for distributed environments. Unfortunately, the originally proposed estimation method has some problems to guarantee the same accuracy over the full cardinality range. Therefore, a couple of variants have been developed to correct the original estimate by empirical means \cite{Heule2013,Rhodes2015,Sanfilippo2014}. 

An estimator for \ac{HLL} sketches, that does not rely on empirical data and that significantly improves the estimation error, is the historic inverse probability estimator \cite{Ting2014, Cohen2015}. It trades memory efficiency for mergeability. The estimator needs to be continuously updated while inserting elements and the estimate depends on the insertion order. Moreover, the estimator cannot be further used after merging two sketches, which limits its application to single data streams. If this restriction is acceptable, the self-learning bitmap \cite{Chen2011}, which provides a similar trade-off and also needs less space than the original \ac{HLL} method, could be used alternatively.

Sometimes not only the number of distinct elements but also a sample of them is needed in order to allow later filtering according to some predicate and estimating the cardinalities of corresponding subsets. In this case the k-minimum values algorithm \cite{Beyer2007, Cohen2007} is the method of choice. It needs more space than the \ac{HLL} algorithm, but also allows set manipulations like construction of intersections, relative complements, or unions \cite{Dasgupta2015}. The latter operation is the only one that is natively supported by \ac{HLL} sketches. A sketch that represents the set operation result is not always needed. One approach to estimate the corresponding result cardinality directly is based on the inclusion-exclusion principle, which however can lead to large errors, especially if the result is small compared to the input set sizes \cite{Dasgupta2015}. Therefore, it was proposed to combine \ac{HLL} sketches with minwise hashing \cite{Pascoe2013, Cohen2016}, which improves the estimation error, even though at the expense of significant more space consumption. 
It was recently pointed out without special focus on \ac{HLL} sketches, that the application of the \ac{ML} method to the joint likelihood function of two probabilistic data structures  leads to better cardinality estimates for intersections \cite{Ting2016}.

\section{HyperLogLog Data Structure}
\label{sec:hyperloglog_data_structure}

\begin{algorithm}[t]
\caption{Insertion of a data element $\symDataItem$ into a \ac{HLL} sketch. All registers $\boldsymbol{\symRegValVariate} = (\symRegValVariate_1,\ldots,\symRegValVariate_\symNumReg)$ start from zero.}
\label{alg:insert}
\begin{algorithmic}
\Procedure {InsertElement}{\symDataItem}
\State $\langle \symBitRepA_1, \ldots, \symBitRepA_\symPrecision,\symBitRepB_1,\ldots,\symBitRepB_\symRegRange\rangle_2 \gets$ $(\symPrecision + \symRegRange)$-bit hash value of $\symDataItem$
\State $\symIndexI \gets 1+ \langle \symBitRepA_1, \ldots, \symBitRepA_\symPrecision\rangle_2$
\State $\symRegVal \gets \min(\{\symS\mid \symBitRepB_\symS = 1\}\cup {\{\symRegRange+1\}} )$
\If{$\symRegVal>\symRegValVariate_\symIndexI$}
\State $\symRegValVariate_\symIndexI \gets\symRegVal$
\EndIf
\EndProcedure
\end{algorithmic}
\end{algorithm}

The \ac{HLL} algorithm collects information of incoming elements into a 
very compact sketching data structure, that finally allows to estimate the number of distinct elements. The data structure consists of $\symNumReg = 2^\symPrecision$ registers. All registers start with zero initial value.
The insertion of a data element into a \ac{HLL} data structure requires the calculation of a $(\symPrecision+\symRegRange)$-bit hash value. The leading $\symPrecision$ bits of the hash value are used to select one of the $2^\symPrecision$ registers. Among the next following $\symRegRange$ bits, the position of the first 1-bit is determined which is a value in the range $[1,\symRegRange+1]$. The value $\symRegRange+1$ is used, if all $\symRegRange$ bits are zeros. If the position of the first 1-bit exceeds the current value of the selected register, the register value is replaced. The complete update procedure is shown in \cref{alg:insert}.

A \ac{HLL} sketch can be characterized by the parameter pair $(\symPrecision, \symRegRange)$ where the precision parameter $\symPrecision$ controls the relative estimation error which scales like $1/\sqrt{\symNumReg}$ \cite{Flajolet2007} while $\symRegRange$ defines the possible range for registers which is $\lbrace 0, 1,\ldots,\symRegRange+1\rbrace$. The case $\symRegRange=0$ corresponds to a bit array and shows that the \ac{HLL} algorithm can be regarded as generalization of linear counting \cite{Whang1990}. The number of consumed hash value bits $\symPrecision+\symRegRange$ defines the maximum cardinality that can be tracked. Obviously, if the cardinality reaches values in the order of $2^{\symPrecision+\symRegRange}$, hash collisions will become more apparent and the estimation error will increase. 

\cref{alg:insert} has some properties which are especially useful for distributed data streams. First, the insertion order of elements has no influence on the final sketch state. Furthermore, any two \ac{HLL} sketches with same parameters $(\symPrecision, \symRegRange)$ representing two different sets can be easily merged. The sketch that represents the union of both sets can be constructed by taking the register-wise maximum values.

The state of a \ac{HLL} sketch is described by the vector $\boldsymbol{\symRegValVariate} = (\symRegValVariate_1,\ldots,\symRegValVariate_\symNumReg)$. Under the assumption of a uniform hash function the inserted elements are distributed over all $\symNumReg$ registers according to a multinomial distribution with equal probabilities $1/\symNumReg$ \cite{Flajolet2007}. 
Therefore, any permutation of $\boldsymbol{\symRegValVariate}$ is equally likely for a given cardinality. Thus, the order of register values $\symRegValVariate_1,\ldots,\symRegValVariate_\symNumReg$ contains no information about the cardinality which makes the multiset $\lbrace\symRegValVariate_1,\ldots,\symRegValVariate_\symNumReg\rbrace$ a sufficient statistic for $\symCardinality$.
Since the values of the multiset are all in the range $[0, \symRegRange+1]$, the multiset can also be written as $\lbrace\symRegValVariate_1,\ldots,\symRegValVariate_\symNumReg\rbrace = 0^{\symCountVariate_0}1^{\symCountVariate_1}\cdots\symRegRange^{\symCountVariate_{\symRegRange}}(\symRegRange+1)^{\symCountVariate_{\symRegRange+1}}$ where $\symCountVariate_\symRegVal$ is the multiplicity of value $\symRegVal$. 
As a consequence, the multiplicity vector $\boldsymbol{\symCountVariate} := (\symCountVariate_0,\ldots,\symCountVariate_{\symRegRange+1})$, which corresponds to the register value histogram, is also a sufficient statistic for the cardinality. By definition we have $\sum_{\symRegVal=0}^{\symRegRange+1}\symCountVariate_{\symRegVal}=\symNumReg$.

\subsection{Poisson Approximation}
\label{sec:poisson_approximation}
The multinomial distribution is the reason that register values are statistically dependent and that further analysis is difficult. For simplification a Poisson model can be used \cite{Flajolet2007}, which assumes that the cardinality itself is distributed according to a Poisson distribution
$\symCardinality \sim \text{Poisson}(\symPoissonRate)$.
Under the Poisson model the register values are independent and identically distributed according to
\begin{equation}
\label{equ:register_value_distribution}
\symProbability(\symRegValVariate \leq \symRegVal\vert\symPoissonRate)
=
\begin{cases}
0 & \symRegVal < 0 \\
e^{-\frac{\symPoissonRate}{\symNumReg 2^\symRegVal}} & 0\leq \symRegVal \leq \symRegRange \\
1 & \symRegVal > \symRegRange.
\end{cases}
\end{equation}

The Poisson approximation makes it easier to find an estimator $\symPoissonRateEstimate = \symPoissonRateEstimate(\boldsymbol{\symRegValVariate})$ for the Poisson rate $\symPoissonRate$ than for the cardinality $\symCardinality$ under the fixed-size model. Depoissonization finally allows to translate the estimates back to the fixed-size model. Assume we have found an estimator $\symPoissonRateEstimate$ for the Poisson rate that is unbiased $\symExpectation(\symPoissonRateEstimate\vert\symPoissonRate) = \symPoissonRate$ for all $\symPoissonRate\geq 0$.
This implies $\symExpectation(\symPoissonRateEstimate\vert\symCardinality) = \symCardinality$ as it is the only solution of 
\begin{equation*}
\symExpectation(\symPoissonRateEstimate\vert\symPoissonRate) 
 = 
\sum_{\symCardinality=0}^\infty \symExpectation(\symPoissonRateEstimate\vert\symCardinality) e^{-\symPoissonRate}\frac{\symPoissonRate^\symCardinality}{\symCardinality!}
= \symPoissonRate
\quad
\text{for all}\ \symPoissonRate\geq 0.
\end{equation*}
Hence, the unbiased estimator $\symPoissonRateEstimate$ conditioned on $\symCardinality$ is also an unbiased estimator for $\symCardinality$, which motivates us to use $\symPoissonRateEstimate$ directly as estimator for the cardinality $\symCardinalityEstimate := \symPoissonRateEstimate$. As our results will show later, this Poisson approximation works well over the full cardinality range, even for estimators that are not exactly unbiased.

\section{Original Estimation Approach}
\label{sec:cardinality_estimation}
The original cardinality estimator \cite{Flajolet2007} is based on the idea that the number of distinct element insertions a register needs to reach the value $\symRegVal$ is proportional to $\symNumReg  2^{\symRegVal}$. Given that, a rough cardinality estimate can be obtained by averaging the values $\lbrace \symNumReg 2^{\symRegValVariate_1},\ldots,\symNumReg2^{\symRegValVariate_\symNumReg}\rbrace$. 
The harmonic mean was found to work best as it is less sensitive to outliers. The result is the so-called raw estimator given by
\begin{equation}
\label{equ:raw_estimator}
\symCardinalityRawEstimate
=
\alpha_\symNumReg
\frac{\symNumReg}
{\frac{1}{\symNumReg 2^{\symRegValVariate_1}}+\ldots+\frac{1}{\symNumReg2^{\symRegValVariate_\symNumReg}}}
= 
\frac{\symAlpha_\symNumReg \symNumReg^2}{\sum_{\symRegVal=0}^{\symRegRange+1}\symCountVariate_\symRegVal 2^{-\symRegVal}}.
\end{equation}
Here $\alpha_\symNumReg$ is a bias correction factor \cite{Flajolet2007} which 
can be well approximated by $\symAlpha_\infty := \lim_{\symNumReg\rightarrow\infty} \symAlpha_\symNumReg = \frac{1}{2\log 2}$ in practice, because the additional bias is negligible compared to the overall estimation error.

To investigate the estimation error of the raw estimator and other estimation approaches discussed in the following, we filled  \num{10000} \ac{HLL} sketches with up to 50 billion unique elements.
To speed up computation we assumed a uniform hash function whose values can be simulated by random numbers. We used the Mersenne Twister random number generator with a state size of \num{19937} bits from the C++ standard library. 

\begin{figure}
\centering
\includegraphics[width=1\columnwidth]{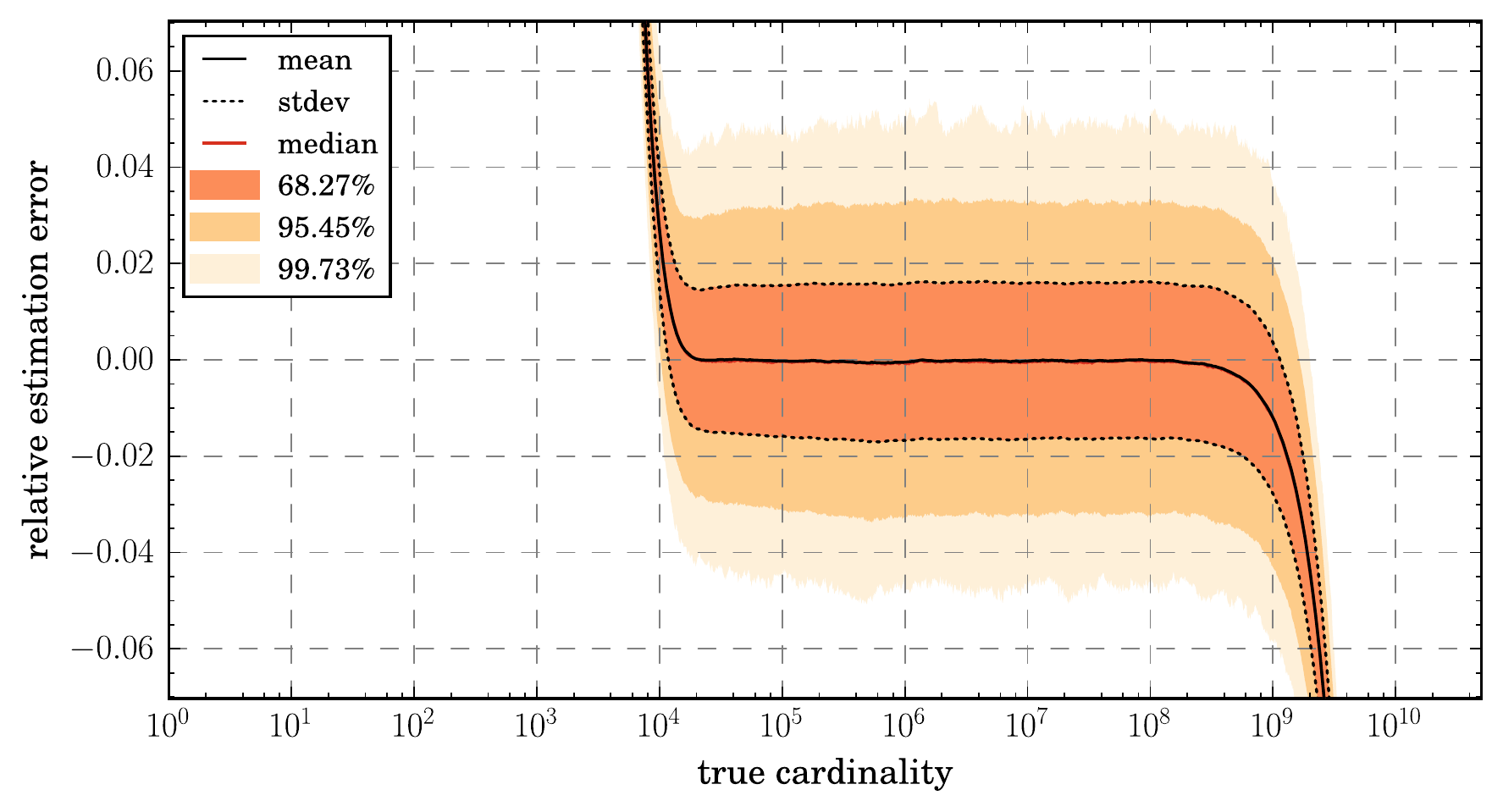}
\caption{Relative error of the raw estimator for \boldmath$\symPrecision = 12$ and $\symRegRange=20$.}
\label{fig:raw_estimate}
\end{figure}

\begin{figure}
\includegraphics[width=1\columnwidth]{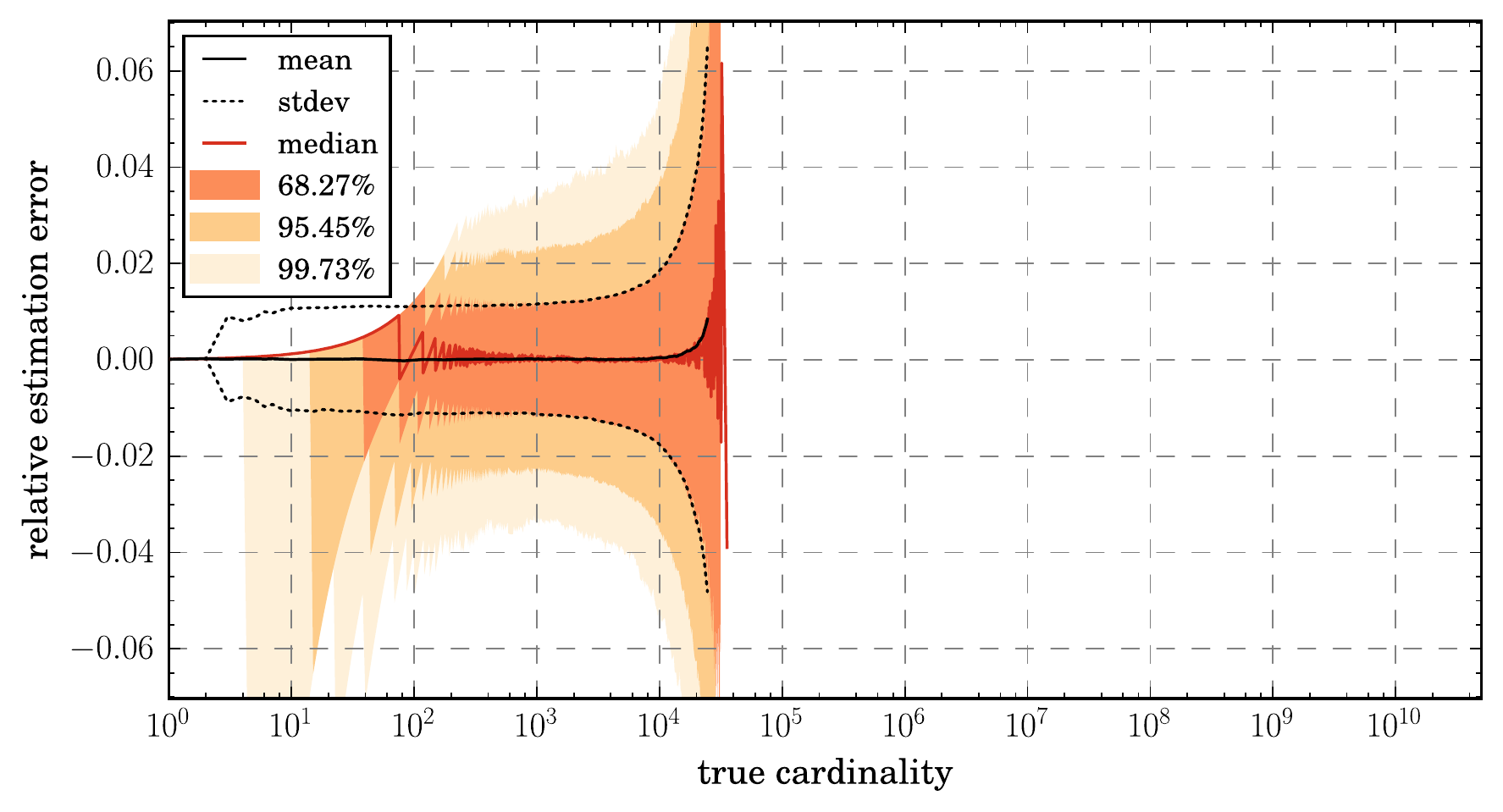}
\caption{Relative error of the linear counting estimator for bitmap size 4096 which corresponds to \boldmath$\symPrecision = 12$ and  $\symRegRange=0$.}
\label{fig:small_range_estimate}
\end{figure}

\cref{fig:raw_estimate} shows the distribution of the relative error of the raw estimator as function of the true cardinality for $\symPrecision=12$ and $\symRegRange=20$. Corrections for small and large cardinalities have been proposed to reduce the obvious bias.
For small cardinalities the \ac{HLL} sketch can be interpreted as bit array by distinguishing between registers with zero and nonzero values. This allows using the linear counting cardinality estimator \cite{Whang1990}
\begin{equation}
\label{equ:linear_counting_estimator}
\symCardinalityEstimate_\text{small} = \symNumReg \log(\symNumReg/\symCountVariate_0).
\end{equation}
The corresponding estimation error is small for small cardinalities and is shown in \cref{fig:small_range_estimate}. It was proposed to use this estimator as long as $\symCardinalityRawEstimate \leq \frac{5}{2}\symNumReg$ where the factor $\frac{5}{2}$ was empirically determined \cite{Flajolet2007}. 
For large cardinalities in the order of $2^{\symPrecision+\symRegRange}$, for which a lot of registers are already in a saturated state, meaning that they have reached the maximum possible value $\symRegRange + 1$, the raw estimator underestimates cardinalities. For the 32-bit hash value case $(\symPrecision+\symRegRange=32)$, which was considered in \cite{Flajolet2007}, following correction formula was proposed  if $\symCardinalityRawEstimate>2^{32}/30\approx\num{1.43e8}$ to take these saturated registers into account
\begin{equation}
\label{equ:large_range_estimate}
\symCardinalityEstimate_\text{large}
=
-2^{32}\log(1-\symCardinalityRawEstimate/2^{32}).
\end{equation}

\begin{figure}
\includegraphics[width=1\columnwidth]{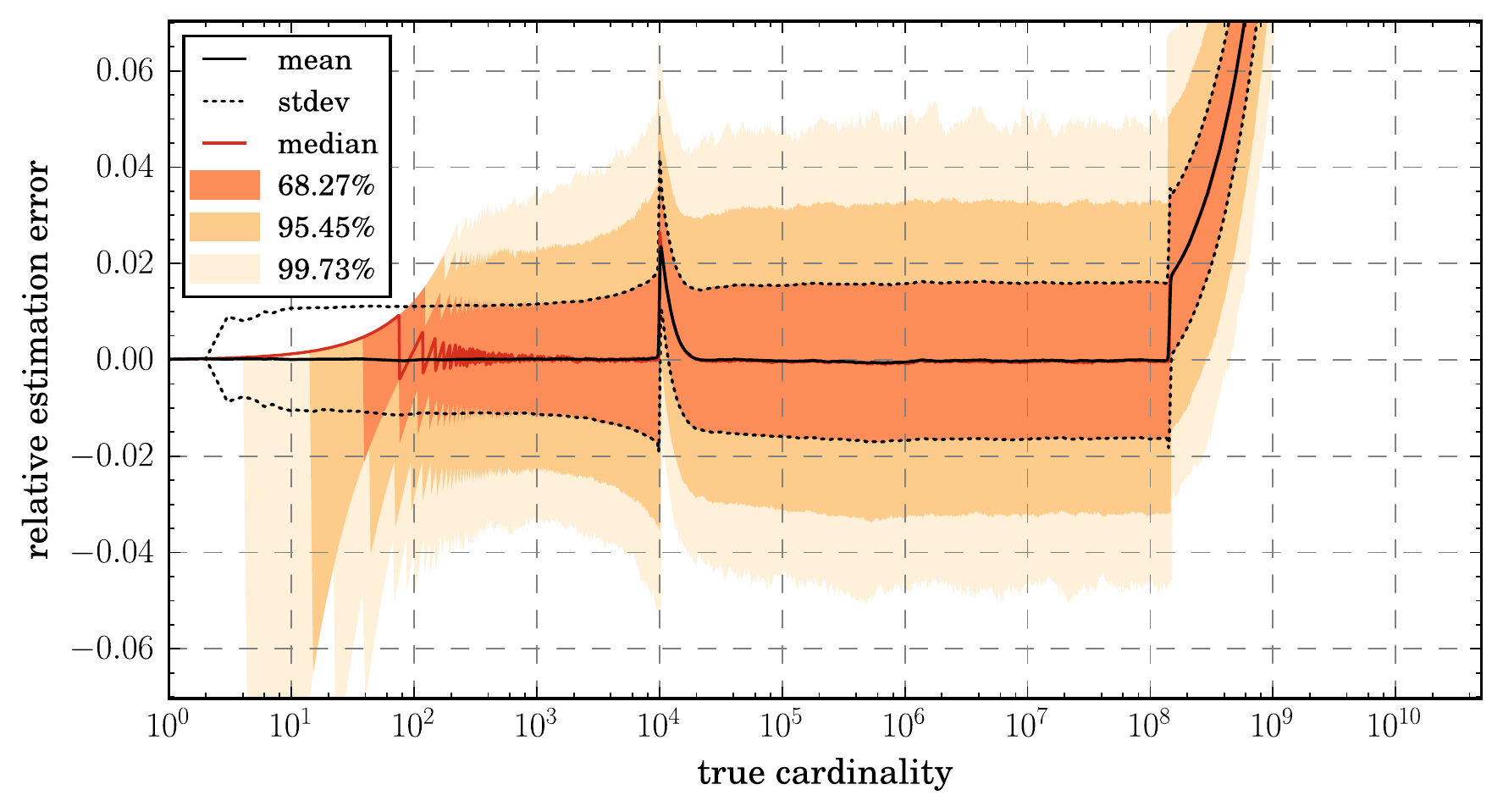}
\caption{Relative estimation error of the original method for \boldmath$\symPrecision = 12$ and $\symRegRange=20$.}
\label{fig:original_estimate}
\end{figure}

The relative estimation error of the original method that includes both corrections is shown in \cref{fig:original_estimate} for the case $\symPrecision=12$ and $\symRegRange=20$. Unfortunately, the ranges where the estimation error is small for $\symCardinalityRawEstimate$ and $\symCardinalityEstimate_\text{small}$ do not overlap, which causes the estimation error to be much larger near the transition region. To reduce the error for cardinalities close to this region, it was proposed to correct the bias of $\symCardinalityRawEstimate$. Empirically collected bias correction data can be either stored as set of interpolation points \cite{Heule2013}, as lookup table \cite{Rhodes2015}, or as best-fitting polynomial \cite{Sanfilippo2014}. However, all these empirical approaches treat the symptom and not the cause.


The large range correction formula \eqref{equ:large_range_estimate} is not satisfying either as it does not reduce the estimation error but makes it even worse. Instead of underestimating cardinalities, they are now overestimated. Another indication for the incorrectness of the proposed large range correction is the fact that it is not even defined for all possible states. For instance, consider a $(\symPrecision,\symRegRange)$-\ac{HLL} sketch with $\symPrecision+\symRegRange=32$ for which all registers are equal to the maximum possible value $\symRegRange+1$. The raw estimate would be $\symCardinalityRawEstimate = \symAlpha_\symNumReg 2^{33}$, which is greater than $2^{32}$ and outside of the domain of the large range correction formula.
A simple approach to avoid the need of any large range corrections is to extend the operating range of the raw estimator to larger cardinalities. This can be easily accomplished by increasing $\symPrecision+\symRegRange$, which corresponds to using hash values with more bits. Each additional bit doubles the operating range which scales like $2^{\symPrecision+\symRegRange}$. However, in case $\symRegRange \geq 31$ the number of possible register values, which are $\lbrace 0, 1, \ldots, \symRegRange+1\rbrace$, exceeds 32 and is no longer representable by 5 bits. Therefore, it was proposed to use 6 bits per register in combination with 64-bit hash values \cite{Heule2013}. Even larger hash values are needless in practice, because it is unrealistic to encounter cardinalities of order $2^{64}$.

\subsection{Derivation of the Raw Estimator}
\label{sec:derivation_raw_estimator}
\begin{figure}
\centering
\includegraphics[width=\columnwidth]{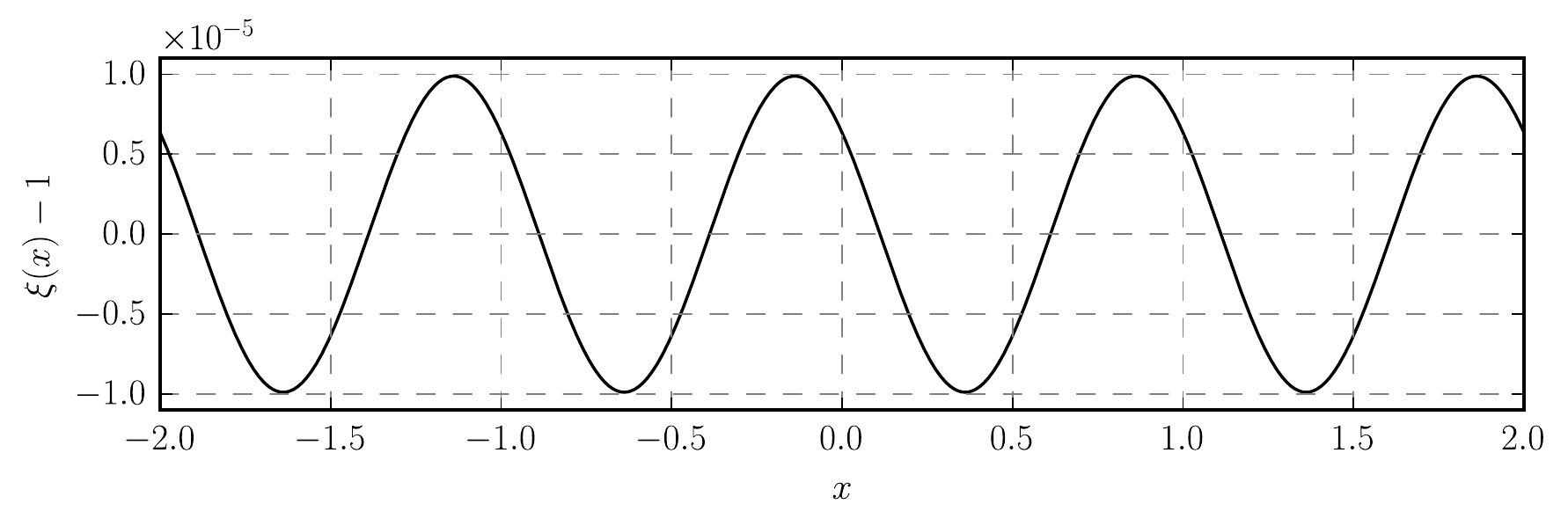}
\caption{The deviation of $\symPowerSeriesFunc(\symX)$ from 1.}
\label{fig:power_series_func}
\end{figure}

To better understand why the raw estimator fails for small and large cardinalities, we start with a brief and simple derivation without the restriction to large cardinalities ($\symCardinality\rightarrow\infty$) and without using complex analysis as in \cite{Flajolet2007}.
Assume that the register values have following cumulative distribution function
\begin{equation}
\label{equ:assumed_register_val_distribution}
\symProbability(\symRegValVariate \leq \symRegVal\vert\symPoissonRate) = e^{-\frac{\symPoissonRate}{\symNumReg 2^{ \symRegVal}}}.
\end{equation}
For now we ignore that this distribution has infinite support and differs from the register value distribution under the Poisson model \eqref{equ:register_value_distribution}, whose support is limited to $[0, \symRegRange+1]$. For a random variable $\symRegValVariate$ obeying \eqref{equ:assumed_register_val_distribution} the expectation of $2^{-\symRegValVariate}$ is given by
\begin{equation}
\label{equ:expectation_power_two}
\symExpectation(2^{-\symRegValVariate})
=
\frac{1}{2}
\sum_{\symRegVal = -\infty}^\infty
2^{-\symRegVal}
e^{-\frac{\symPoissonRate}{\symNumReg 2^{\symRegVal}} }
=
\frac{\symAlpha_\infty\,\symNumReg\,\symPowerSeriesFunc\!\left(\log_2\!\left(\symPoissonRate/\symNumReg\right)\right)}{\symPoissonRate},
\end{equation}
where the function 
\begin{equation*}
\symPowerSeriesFunc(\symX):= \log(2) \sum_{\symRegVal = -\infty}^\infty
2^{\symRegVal+\symX}
e^{-2^{\symRegVal+\symX}}
\end{equation*}
is a smooth periodic oscillating function with mean 1 and an amplitude that can be bounded by $\symEpsPowerSeriesFunc:=\num{9.885e-6}$ as shown in \cref{fig:power_series_func}. This limit can also be found using Fourier analysis \cite{Ertl2017}.
For a large ($\symNumReg\rightarrow \infty$) sample $\symRegValVariate_1,\ldots,\symRegValVariate_\symNumReg$, that is distributed according to \eqref{equ:assumed_register_val_distribution}, we asymptotically have
\begin{equation*}
\symExpectation\!\left(
\frac{1}{2^
{-\symRegValVariate_1}\!+\ldots+2^{-\symRegValVariate_\symNumReg}}
\right)
\underset{\symNumReg\rightarrow \infty}{=}
\frac{1}{\symExpectation
(2^{-\symRegValVariate_1}+\ldots+2^{-\symRegValVariate_\symNumReg})}
=
\frac{1}{\symNumReg \symExpectation(2^{-\symRegValVariate})}.
\end{equation*}
Together with \eqref{equ:expectation_power_two} we obtain
\begin{equation*}
\symPoissonRate
=
\symExpectation\left(
\frac{\symAlpha_\infty\,\symNumReg^2\,\symPowerSeriesFunc(\log_2( \symPoissonRate/\symNumReg))}{2^{-\symRegValVariate_1}+\ldots+2^{-\symRegValVariate_\symNumReg}}
\right)
\quad
\text{for $\symNumReg\rightarrow \infty$}.
\end{equation*}
Therefore, the asymptotic relative bias of 
\begin{equation*}
\symPoissonRateEstimate 
= 
\frac{\symAlpha_\infty\,\symNumReg^2}{2^{-\symRegValVariate_1}+\ldots+2^{-\symRegValVariate_\symNumReg}}
\end{equation*}
is bounded by $\symEpsPowerSeriesFunc$, which makes this statistic a good estimator for the Poisson parameter. It also corresponds to the raw estimator \eqref{equ:raw_estimator}, if the Poisson parameter estimate is used as cardinality estimate as discussed in \cref{sec:poisson_approximation}.

\subsection{Limitations of the Raw Estimator}
The raw estimator is based on two prerequisites. In practice, only the first requiring $\symNumReg$ to be sufficiently large is satisfied. However, the second assuming that the distribution of register values \eqref{equ:register_value_distribution} can be approximated by \eqref{equ:assumed_register_val_distribution} is not always true.
A random variable $\symRegValVariate'$ with cumulative distribution function \eqref{equ:assumed_register_val_distribution} can be transformed into a random variable $\symRegValVariate$ with cumulative distribution function \eqref{equ:register_value_distribution} using
\begin{equation}
\label{equ:dist_transformation}
\symRegValVariate = \min\!\left(\max\!\left(\symRegValVariate',0\right), \symRegRange+1\right).
\end{equation}
Therefore, register values $\symRegValVariate_1,\ldots,\symRegValVariate_\symNumReg$ can be seen as the result after applying this transformation to a sample $\symRegValVariate'_1,\ldots,\symRegValVariate'_\symNumReg$ from \eqref{equ:assumed_register_val_distribution}. If all registers values are in the range $[1,\symRegRange]$, they must be identical to the values $\symRegValVariate'_1,\ldots,\symRegValVariate'_\symNumReg$. In other words, the observed register values are also a plausible sample of distribution \eqref{equ:assumed_register_val_distribution}. Hence, as long as all or at least most register values are in the range $[1,\symRegRange]$, which is the case if $2^\symPrecision \ll \symPoissonRate \ll 2^{\symPrecision
+\symRegRange}$, the approximation of \eqref{equ:register_value_distribution} by \eqref{equ:assumed_register_val_distribution} is valid. This explains why the raw estimator works best for intermediate cardinalities. However, for small and large cardinalities many register values are equal to 0 or $\symRegRange+1$, respectively, which contradicts \eqref{equ:assumed_register_val_distribution} and ends up in the observed bias.

\section{Improved Estimator}
If we knew the values $\symRegValVariate'_1,\ldots,\symRegValVariate'_\symNumReg$ for which transformation \eqref{equ:dist_transformation} led to the observed register values $\symRegValVariate_1,\ldots,\symRegValVariate_\symNumReg$, we would be able to use the raw estimator
\begin{equation}
\label{equ:raw_estimate_with_multiplicities}
\symPoissonRateEstimate 
= 
\frac{\symAlpha_\infty\,\symNumReg^2}{\sum_{\symRegVal=-\infty}^\infty \symCountVariate'_\symRegVal 2^{-\symRegVal}}
\end{equation}
where $\symCountVariate'_\symRegVal := \vert\lbrace \symIndexI\vert \symRegVal = \symRegValVariate'_\symIndexI\rbrace\vert$ are the multiplicities of value $\symRegVal$ in $\lbrace\symRegValVariate'_1,\ldots,\symRegValVariate'_\symNumReg\rbrace$. Due to \eqref{equ:dist_transformation}, the multiplicities $\symCountVariate'_\symRegVal$ and the multiplicities $\symCountVariate_\symRegVal$ for the observed register values have following relationships
\begin{equation}
\label{equ:multiplicity_transformation}
\symCountVariate_0 = \textstyle\sum_{\symRegVal = -\infty}^{0} \symCountVariate'_\symRegVal,\
\symCountVariate_\symRegVal =  \symCountVariate'_\symRegVal\ (1\leq\symRegVal\leq\symRegRange),\
\symCountVariate_{\symRegRange+1} = \textstyle\sum_{\symRegVal = \symRegRange + 1}^{\infty} \symCountVariate'_\symRegVal.
\end{equation}
The idea is now to find estimates $\hat{\symCount}'_\symRegVal$ for all $\symRegVal\in\mathbb{Z}$ and use them as replacements for $\symCountVariate'_\symRegVal$ in \eqref{equ:raw_estimate_with_multiplicities}. For $\symRegVal\in[1,\symRegRange]$ we can use the trivial estimators 
$\hat{\symCount}'_\symRegVal := \symCountVariate_\symRegVal$.
Estimators for $\symRegVal\leq 0$ and $\symRegVal\geq\symRegRange+1$ can be found by considering the expectation of $\symCountVariate'_\symRegVal$
\begin{equation*}
\symExpectation(\symCountVariate'_\symRegVal)
=
\symNumReg\,
\symProbability(\symRegValVariate' = \symRegVal\vert\symPoissonRate)
=
\symNumReg\,e^{-\frac{\symPoissonRate}{\symNumReg 2^{\symRegVal}}}\left(1-e^{-\frac{\symPoissonRate}{\symNumReg 2^{\symRegVal}}}\right).
\end{equation*}
According to \eqref{equ:register_value_distribution} we have $\symExpectation(\symCountVariate_0/\symNumReg)=
e^{-\frac{\symPoissonRate}{\symNumReg}}$ and $\symExpectation(1-\symCountVariate_{\symRegRange+1}/\symNumReg)=
e^{-\frac{\symPoissonRate}{\symNumReg 2^\symRegRange}}$ which gives 
\begin{align*}
\symExpectation(\symCountVariate'_\symRegVal)
&=
\symNumReg\,
(\symExpectation(\symCountVariate_0/\symNumReg))^{2^{-\symRegVal}}
\left(1-\left(\symExpectation(\symCountVariate_0/\symNumReg)\right)^{2^{-\symRegVal}}\right),
\\
&=
\symNumReg\,
(\symExpectation(1-\symCountVariate_{\symRegRange+1}/\symNumReg))^{2^{\symRegRange-\symRegVal}}
\left(1-(\symExpectation(1-\symCountVariate_{\symRegRange+1}/\symNumReg))^{2^{\symRegRange-\symRegVal}}\right).
\end{align*}
These two expressions for the expectation suggest to use
\begin{align*}
\hat{\symCount}'_\symRegVal
&=
\symNumReg\,
(\symCountVariate_0/\symNumReg)^{2^{-\symRegVal}}
\left(1-(\symCountVariate_0/\symNumReg)^{2^{-\symRegVal}}\right)\quad\text{and}
\\
\hat{\symCount}'_\symRegVal
&=
\symNumReg\,
(1-\symCountVariate_{\symRegRange+1}/\symNumReg)^{2^{\symRegRange-\symRegVal}}
\left(1-(1-\symCountVariate_{\symRegRange+1}/\symNumReg)^{2^{\symRegRange-\symRegVal}}\right)
\end{align*}
as estimators for $\symRegVal \leq 0$ and $\symRegVal \geq \symRegRange+1$, respectively.
Both estimators conserve the mass of zero-valued and saturated registers, because  
\eqref{equ:multiplicity_transformation} is satisfied,
if $\symCountVariate'_\symRegVal$ is replaced by $\hat{\symCount}'_\symRegVal$. 
Plugging all these estimators into \eqref{equ:raw_estimate_with_multiplicities} as replacements for $\symCountVariate'_\symRegVal$ finally gives 
\begin{equation}
\label{equ:correctedestimator}
\symPoissonRateEstimate 
= 
\frac{\symAlpha_\infty\symNumReg^2}
{
\symNumReg\,\symSmallCorrectionFunc(\symCountVariate_0/\symNumReg) + \sum_{\symRegVal=1}^\symRegRange \symCountVariate_\symRegVal 2^{-\symRegVal} + \symNumReg\, \symLargeCorrectionFunc(1-\symCountVariate_{\symRegRange+1}/\symNumReg) 2^{-\symRegRange}
}
\end{equation}
which we call the improved estimator. Here $\symNumReg\,\symSmallCorrectionFunc(\symCountVariate_0/\symNumReg)$ and $2\symNumReg\,\symLargeCorrectionFunc(1-\symCountVariate_{\symRegRange+1}/\symNumReg)$ are replacements for  $\symCountVariate_0$ and $\symCountVariate_{\symRegRange+1}$ in the raw estimator \eqref{equ:raw_estimator}, respectively. The functions $\symSmallCorrectionFunc$ and $\symLargeCorrectionFunc$ are defined as
\begin{align}
\label{equ:sigma}
\symSmallCorrectionFunc(\symX) &:= 
\symX
+
\sum_{\symRegVal=1}^\infty
\symX^{2^\symRegVal} 2^{\symRegVal-1},
\\
\label{equ:tau}
\symLargeCorrectionFunc(\symX)
&:=
\frac{1}{3}
\left(
1-\symX
-
\sum_{\symRegVal=1}^\infty
\left(
1-
\symX^{2^{-\symRegVal}}
\right)^{\!2}
2^{-\symRegVal}
\right).
\end{align}
We can cross-check the new estimator for the linear counting case with $\symRegRange=0$. Using the identity 
\begin{equation}
\label{equ:sigma_tau_relationship}
\symSmallCorrectionFunc(\symX) + \symLargeCorrectionFunc(\symX) = \symAlpha_\infty\symPowerSeriesFunc(\log_2(\log(1/\symX)))/\log(1/\symX),
\end{equation}
 we get
\begin{equation*}
\symPoissonRateEstimate 
= 
\frac{\symAlpha_\infty\symNumReg}{
\symSmallCorrectionFunc(\symCountVariate_0/\symNumReg)
+
\symLargeCorrectionFunc(\symCountVariate_0/\symNumReg)
}
=
\frac{\symNumReg\log\!\left(\symNumReg/\symCountVariate_0\right)}{\symPowerSeriesFunc\!\left(\log_2\!\left(\log\!\left(\symNumReg/\symCountVariate_0\right)\right)\right)}
\end{equation*}
which is as expected almost identical to the linear counting estimator \eqref{equ:linear_counting_estimator}, because $\symPowerSeriesFunc(\symX)\approx 1$.

The new estimator can be directly translated into an estimation algorithm that does not depend on magic numbers or special cases as previous approaches. Since $\symCountVariate_\symRegVal\in\lbrace0,\ldots,\symNumReg\rbrace$, the values for $\symNumReg\,\symSmallCorrectionFunc(\symCountVariate_0/\symNumReg)$ and $\symNumReg\, \symLargeCorrectionFunc(1-\symCountVariate_{\symRegRange+1}/\symNumReg)$ can be precalculated and kept in lookup tables of size $\symNumReg+1$. In this way a complete branch-free cardinality estimation can be realized. However, on-demand calculation of $\symSmallCorrectionFunc$ and $\symLargeCorrectionFunc$ is very fast as well. The series \eqref{equ:sigma} converges quadratically for all $\symX\in[0,1)$ and its terms can be calculated recursively using elementary operations. The case $\symX=1$ needs special handling, because the series diverges and causes an infinite denominator in \eqref{equ:correctedestimator} and therefore a vanishing cardinality estimate. As this case only occurs if all register values are in initial state ($\symCountVariate_0=\symNumReg$), this is exactly what is expected. 
Apart from the trivial roots at $\symX=0$ and $\symX=1$, the calculation of $\symLargeCorrectionFunc$ is slightly more expensive, because it involves square root evaluations and its series converges only linearly. Since $1-\symX^{2^{-\symRegVal}}\leq -\log(\symX) 2^{-\symRegVal}$ for $\symX\in(0,1]$ its convergence speed is comparable to a geometric series with ratio $1/8$. It is also thinkable to calculate $\symLargeCorrectionFunc$ using the approximation $\symLargeCorrectionFunc(\symX)\approx \symAlpha_\infty/\log(1/\symX)-\symSmallCorrectionFunc(\symX)$ which can be obtained from \eqref{equ:sigma_tau_relationship} and $\symPowerSeriesFunc(\symX)\approx 1$. The advantage is that the calculation of $\symSmallCorrectionFunc$ and the additional logarithm is slightly faster than the direct calculation of $\symLargeCorrectionFunc$ using \eqref{equ:tau}. 
However, for arguments close to 1 both terms may become very large compared to their difference, which requires special care to avoid numerical cancellation. 
The calculation of $\symLargeCorrectionFunc$ can be omitted at all, if the \ac{HLL} parameters are chosen such that 
$2^{\symPrecision+\symRegRange}$ is much larger than the
expected cardinality. The number of saturated registers is negligible in this case ($\symCountVariate_{\symRegRange+1}\approx 0$) and therefore $\symLargeCorrectionFunc(1 - \symCountVariate_{\symRegRange+1}/\symNumReg)\approx\symLargeCorrectionFunc(1) = 0$.

\subsection{Results}

\begin{figure}
\centering
\includegraphics[width=1\columnwidth]{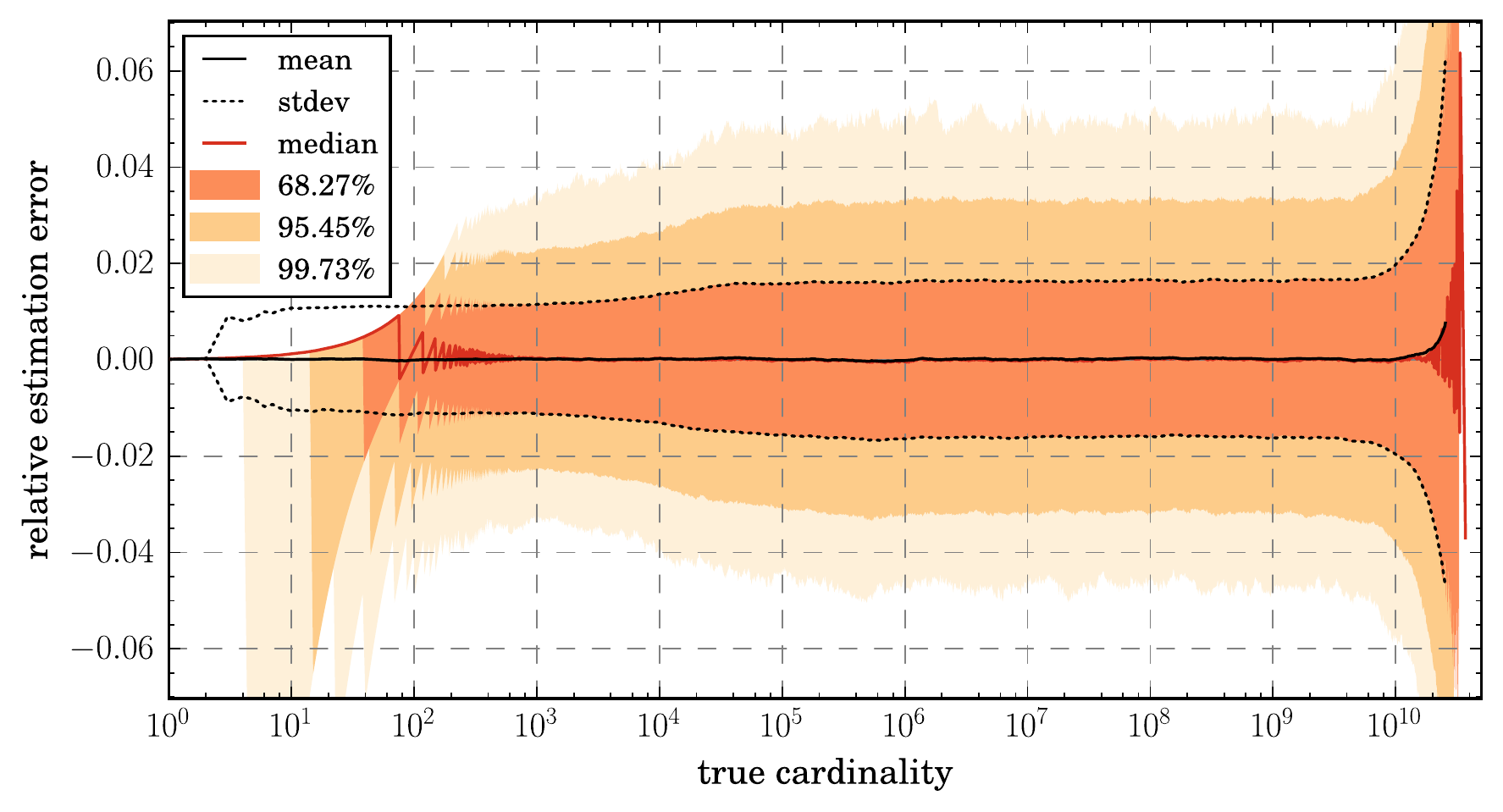}
\caption{Relative error of the improved estimator for \boldmath$\symPrecision = 12$ and $\symRegRange=20$.}
\label{fig:raw_corrected_estimation_error_12_20}
\end{figure}

\begin{figure}
\centering
\includegraphics[width=1\columnwidth]{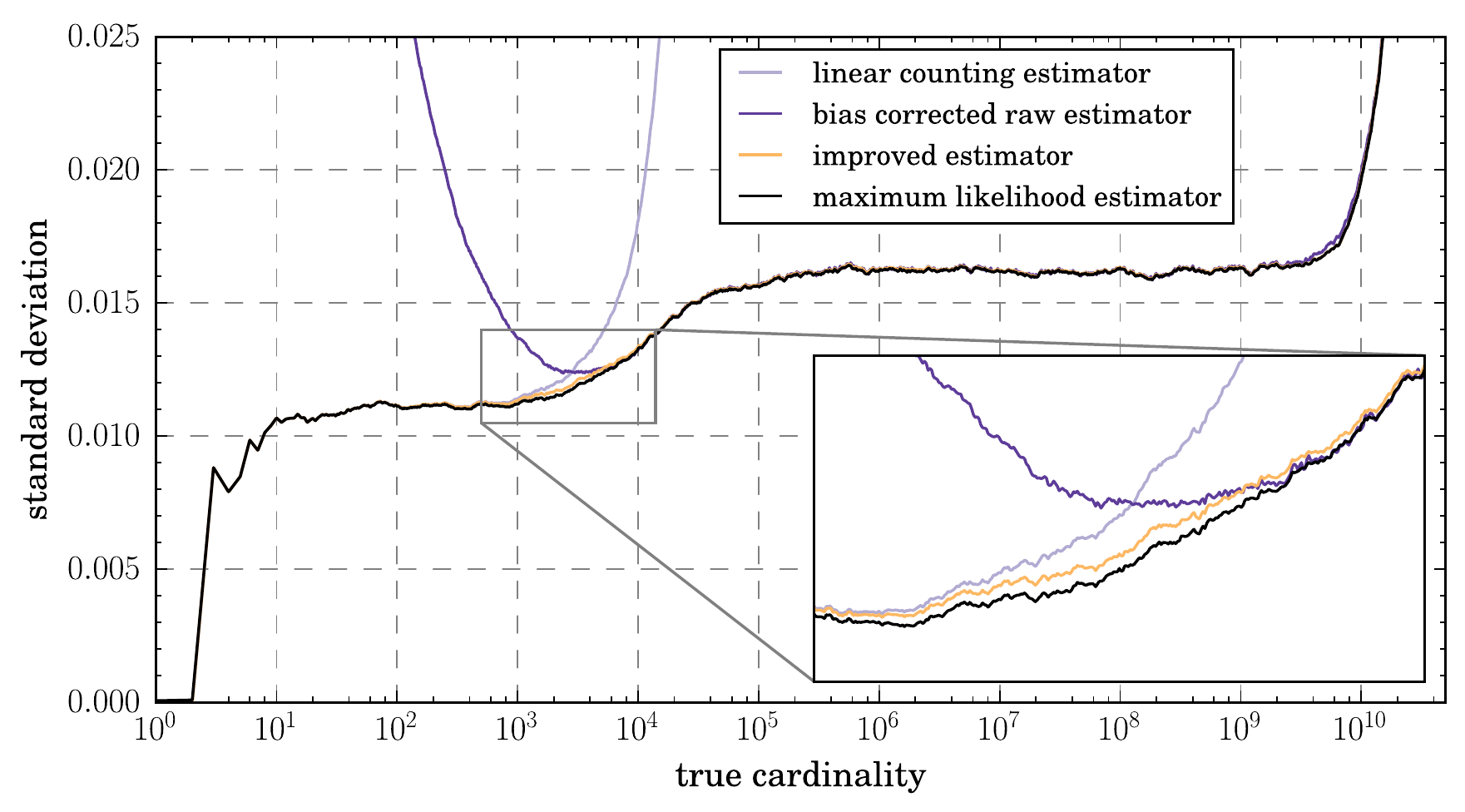}
\caption{Standard deviations of the relative error of different cardinality estimators for \boldmath$\symPrecision = 12$ and $\symRegRange=20$.}
\label{fig:stdev_comparison}
\end{figure}

\label{sec:corrected_raw_estimation_error}
\cref{fig:raw_corrected_estimation_error_12_20} shows the relative estimation error of the new improved estimator, again based on \num{10000} randomly generated \ac{HLL} sketches with parameters $\symPrecision=12$ and $\symRegRange=20$. The experimental results show that  new estimator is unbiased up to cardinalities of 10 billions which is a clear improvement over the raw estimator (compare \cref{fig:raw_estimate}). 
The improved estimator beats the precision of existing methods that apply bias correction on the raw estimator \eqref{equ:raw_estimator} \cite{Heule2013,Rhodes2015,Sanfilippo2014}. Based on the simulated data we have empirically determined the bias correction function $\symBiasCorrectionFunc$ that satisfies $\symCardinality = \symExpectation(\symBiasCorrectionFunc(\symCardinalityRawEstimate)\vert\symCardinality)$ for all cardinalities. By definition, 
the estimator $\symCardinalityRawEstimate':=\symBiasCorrectionFunc(\symCardinalityRawEstimate)$ is unbiased and a function of the raw estimator. Its standard deviation is compared with that of the improved estimator in \cref{fig:stdev_comparison}. For cardinalities smaller than \num{e4} the empirical bias correction approach is not very precise. This is the reason why all previous approaches had to switch over to the linear counting estimator at some point. The standard deviation of the linear counting estimator is also shown in \cref{fig:stdev_comparison}. Obviously, the previous approaches cannot do better than given by the minimum of both curves for linear counting and raw estimator. In practice, the standard deviation of the combined approach is even larger, because the choice between both estimators must be made based on an estimate and not on the true cardinality, for which the intersection point of both curves, which is approximately \num{3e3} in \cref{fig:stdev_comparison}, would be the ideal transition point. In contrast, the improved estimator performs well over the entire cardinality range. 

\begin{figure}
\centering
\includegraphics[width=1\columnwidth]{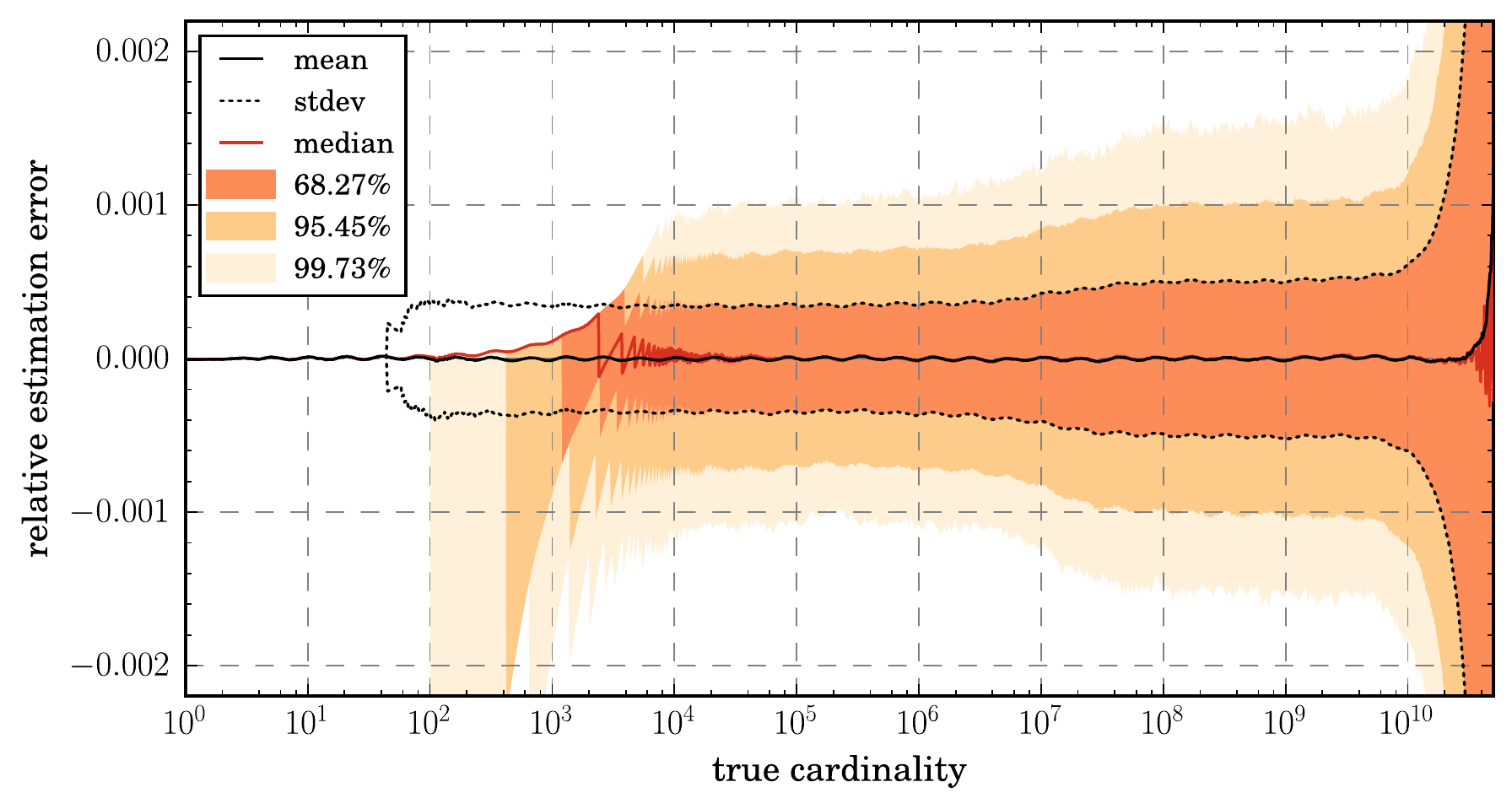}
\caption{Relative error of the improved estimator for \boldmath$\symPrecision = 22$ and $\symRegRange=10$.}
\label{fig:raw_corrected_estimation_error_22_10}
\end{figure}

We also investigated the estimation error of the improved estimator for the case $\symPrecision=22$, $\symRegRange=10$ as shown in \cref{fig:raw_corrected_estimation_error_22_10}. 
The standard deviation is by a factor 32 smaller according to the $1/\sqrt{\symNumReg}$ error scaling law. Again, cardinalities up to order $2^{\symPrecision+\symRegRange}\approx\num{4e9}$ can be well estimated, because $\symPrecision+\symRegRange=32$ was kept the same for both investigated \ac{HLL} configurations. For $\symPrecision=22$ a small oscillating bias becomes apparent, which is caused by approximating the periodic function $\symPowerSeriesFunc$ by a constant (see \cref{sec:derivation_raw_estimator}).

\section{Maximum Likelihood Estimation}
\label{sec:max_likelihood_estimation}
We know from \cref{sec:poisson_approximation} that any unbiased estimator for the Poisson parameter is also an unbiased estimator for the cardinality. Moreover, we know that under suitable regularity conditions of the probability mass function the \ac{ML} estimator is asymptotically efficient \cite{Casella2002}. This means, if the number of registers $\symNumReg$ is sufficiently large, the estimator should be unbiased.

The log-likelihood function for given register values $\boldsymbol{\symRegValVariate}$, which are assumed to be distributed according to \eqref{equ:register_value_distribution} under the Poisson model, is
\begin{equation}
\label{equ:log_likelihood_single}
\log \mathcal{\symLikelihood}(\symPoissonRate\vert\boldsymbol{\symRegValVariate}) = 
\sum_{\symRegVal=1}^{\symRegRange+1}
\log\left(1-e^{-\frac{\symPoissonRate}{\symNumReg 2^{\min(\symRegVal,\symRegRange)}}}\right)
\symCountVariate_\symRegVal
-
\frac{\symPoissonRate}{\symNumReg}\sum_{\symRegVal=0}^{\symRegRange}\frac{\symCountVariate_\symRegVal}{2^\symRegVal}
.
\end{equation}
After differentiation and multiplying by $\symPoissonRate$ we find that the \ac{ML} estimate $\symPoissonRateEstimate$ is the unique root of the monotone decreasing function
\begin{equation*}
\symFunc(\symPoissonRate)
=
\sum_{\symRegVal=1}^{\symRegRange+1} \frac{\frac{\symPoissonRate}{\symNumReg 2^{\min(\symRegVal,\symRegRange)}}}{e^{\frac{\symPoissonRate}{\symNumReg 2^{\min(\symRegVal,\symRegRange)}}}-1}
\symCountVariate_\symRegVal
-
\frac{\symPoissonRate}{\symNumReg}\sum_{\symRegVal=0}^\symRegRange \frac{\symCountVariate_\symRegVal}{2^\symRegVal}.
\end{equation*}
Since $\symFunc(0)=\symNumReg-\symCountVariate_0\geq 0$ and $\symFunc$ is at least linear decreasing as long  as $\symCountVariate_{\symRegRange+1} < \symNumReg$, there exists a unique root. The special case $\symCountVariate_{\symRegRange+1} = \symNumReg$, for which all registers have reached the maximum possible value, the \ac{ML} estimate would be positive infinite.

Using $\symFunc(\symPoissonRateEstimate)=0$ and $1-\frac{\symX}{2}\leq \frac{\symX}{e^\symX-1}\leq 1$ we obtain following bounds for $\symPoissonRateEstimate$
\begin{equation}
\label{equ:inequality}
\frac{\symNumReg(\symNumReg-\symCountVariate_0)}
{\symCountVariate_0+\frac{3}{2}\sum_{\symRegVal=1}^{\symRegRange}\frac{\symCountVariate_\symRegVal}{2^\symRegVal} + \frac{\symCountVariate_{\symRegRange+1}}{2^{\symRegRange+1}}}
\leq
\symPoissonRateEstimate
\leq
\frac{\symNumReg (\symNumReg-\symCountVariate_0)}
{\sum_{\symRegVal=0}^{\symRegRange}
\frac{\symCountVariate_\symRegVal}{2^\symRegVal}}.
\end{equation}
If the cardinality is in the intermediate range, where $\symCountVariate_0=\symCountVariate_{\symRegRange+1}=0$ the lower and the upper bound differ only by a constant factor and both are proportional to the harmonic mean of $\lbrace \symNumReg 2^{\symRegValVariate_1},\ldots,\symNumReg 2^{\symRegValVariate_\symNumReg} \rbrace$. Hence, consequent application of the \ac{ML} method would have directly suggested to use a cardinality estimator that is proportional to the harmonic mean without knowing the raw estimator \eqref{equ:raw_estimator} in advance. The history of the \ac{HLL} algorithm shows that the raw estimator was first found after several attempts using the geometric mean \cite{Flajolet2007, Durand2003}.

For $\symRegRange=0$, which corresponds to the linear counting case, the \ac{ML} estimator can be found by analytical means. The result is exactly the linear counting estimator \eqref{equ:linear_counting_estimator} which makes us optimistic that the \ac{ML} method in combination with the Poisson model also works well for the more general \ac{HLL} case $\symRegRange>0$. Since $f$ is convex, Newton-Raphson iteration and the secant method \cite{Press2007} will both converge to the root, provided that the function is positive for the chosen starting points. Even though the secant method has the disadvantage of slower convergence, a single iteration is simpler to calculate as it does not require the evaluation of the first derivative. Possible starting points are zero and the lower bound in \eqref{equ:inequality}. The iteration can be stopped, if the relative increment is below a certain limit $\symStopDelta$. Since the expected estimation error scales according to $1/\sqrt{\symNumReg}$, it makes sense to choose $\symStopDelta = \symStopEpsilon/\sqrt{\symNumReg}$ with some constant $\symStopEpsilon$. For our results presented below we used $\symStopEpsilon = 10^{-2}$. In practice, only a handful of iterations are necessary to satisfy the stop criterion.

\subsection{Results}
\label{sec:maximum_likelihood_estimation_error}

\begin{figure}
\centering
\includegraphics[width=1\columnwidth]{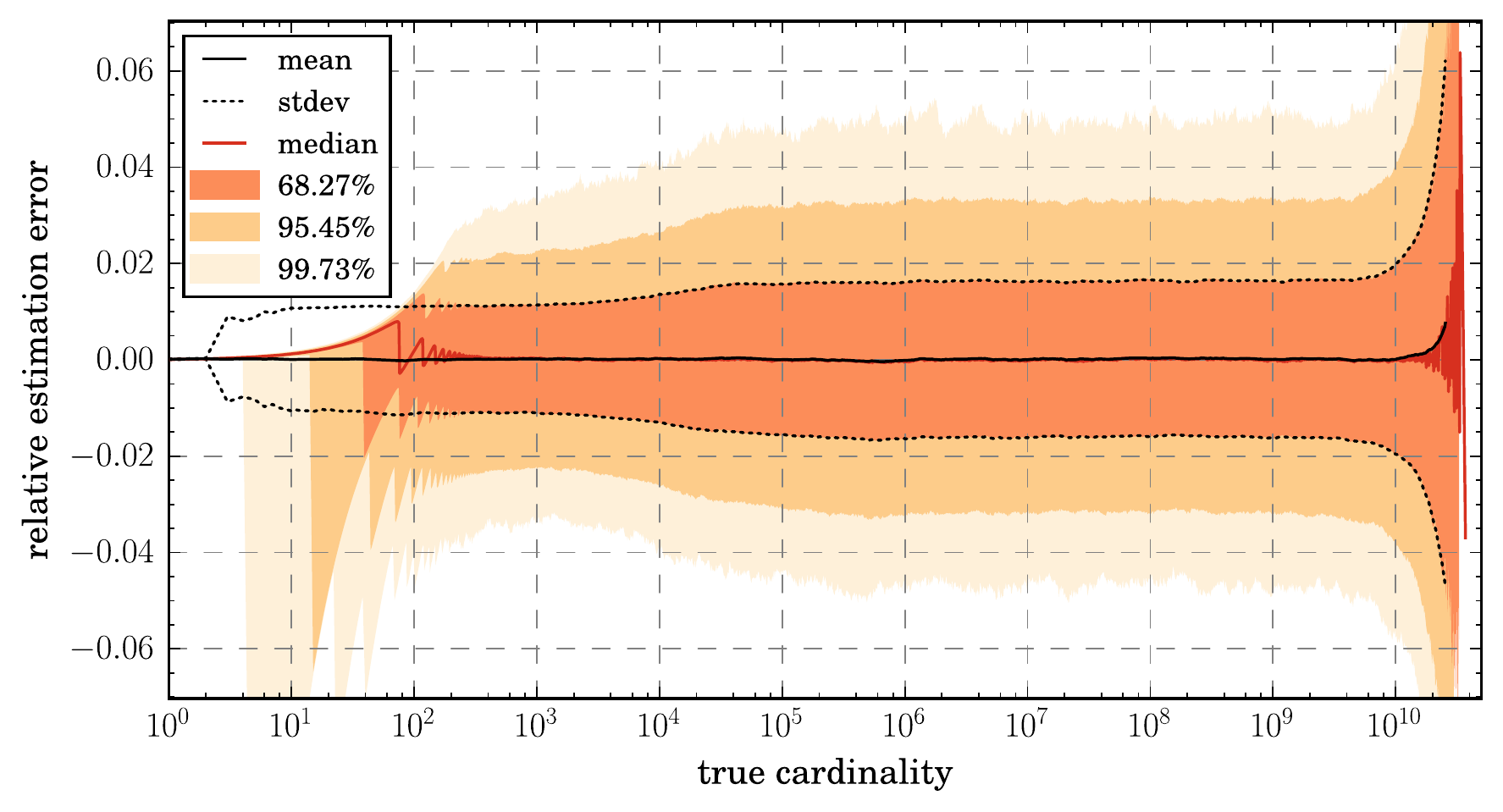}
\caption{Relative error of the \ac{ML} estimator for \boldmath$\symPrecision = 12$ and $\symRegRange=20$.}
\label{fig:max_likelihood_estimation_error_12_20}
\end{figure}
\begin{figure}
\centering
\includegraphics[width=1\columnwidth]{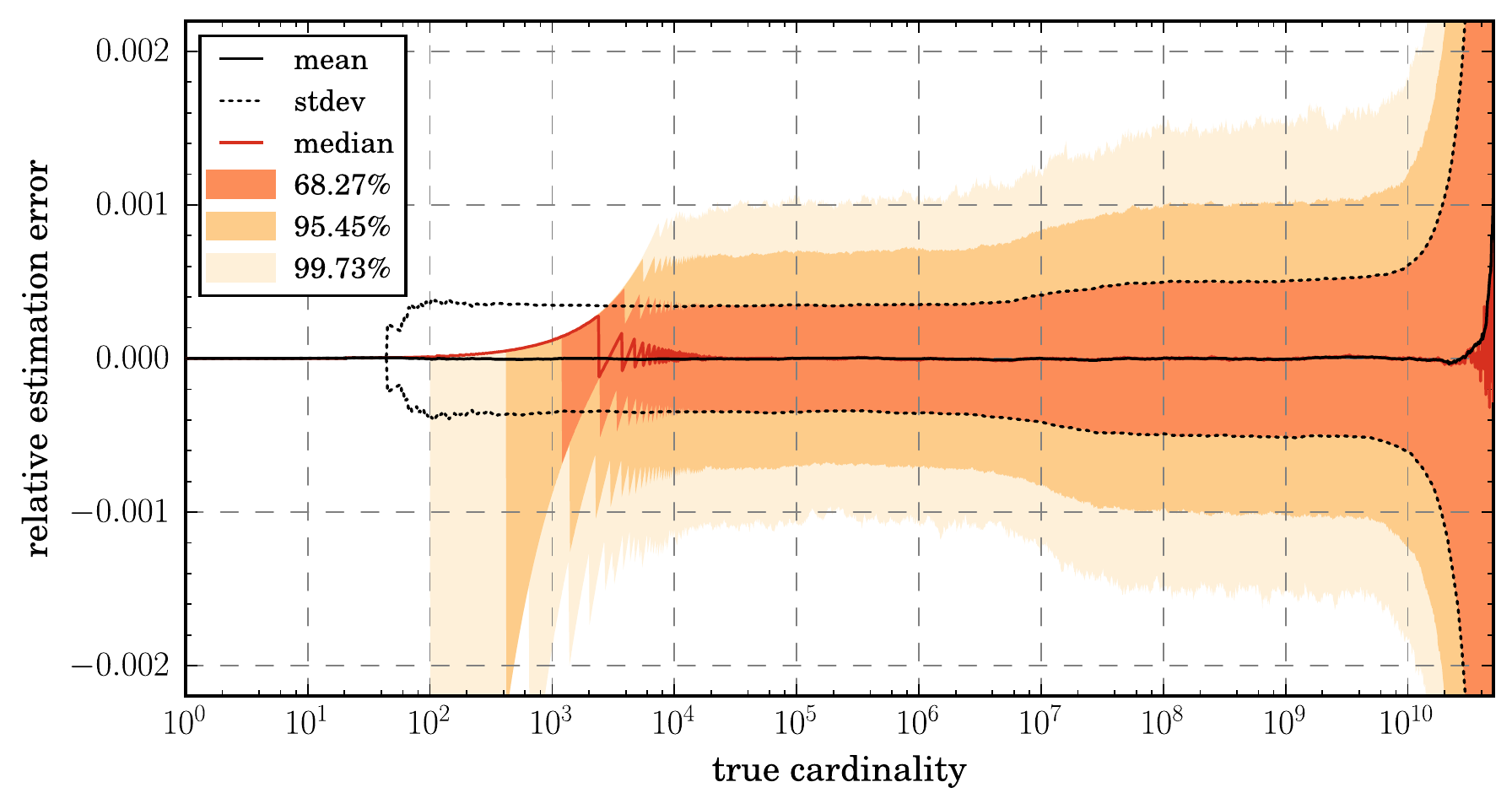}
\caption{Relative error of the \ac{ML} estimator for \boldmath$\symPrecision = 22$ and $\symRegRange=10$.}
\label{fig:max_likelihood_estimation_error_22_10}
\end{figure}

We have investigated the estimation error of the \ac{ML} estimator for both \ac{HLL} configurations as for the improved estimator. 
\cref{fig:max_likelihood_estimation_error_12_20,fig:max_likelihood_estimation_error_22_10} look very similar to 
\cref{fig:raw_corrected_estimation_error_12_20,fig:raw_corrected_estimation_error_22_10}, respectively. For $\symPrecision=12$, $\symRegRange=20$ the \ac{ML} estimator has a somewhat smaller median bias for cardinalities around \num{200}. In addition, the standard deviation of the relative error is marginally better than that of the improved estimator for cardinalities between \num{e3} and \num{e4} as shown in \cref{fig:stdev_comparison}. 
For $\symPrecision=22$, $\symPrecision=10$ we see that the 
the \ac{ML} estimator does not have the small oscillating bias  as the improved estimator. Since all observed improvements  are not very relevant in practice, the improved estimator should be preferred over the \ac{ML} method, because it leads to a simpler and faster algorithm.

\section{Joint Estimation}
\label{sec:cardinality_estimation_set_intersections}

While the union of two sets that are represented by \ac{HLL} sketches can be straightforwardly obtained by taking the register-wise maximums, the computation of cardinalities for other set operations like intersections and relative complements is more challenging. The conventional approach uses the inclusion-exclusion principle 
\begin{equation}
\label{equ:conventional_approach}
\begin{aligned}
&\left\vert\symSetS_1 \setminus \symSetS_2\right\vert
=
\left\vert\symSetS_1 \cup \symSetS_2\right\vert
-
\left\vert\symSetS_2\right\vert,
\quad
\left\vert\symSetS_2 \setminus \symSetS_1\right\vert
=
\left\vert\symSetS_1 \cup \symSetS_2\right\vert
-
\left\vert\symSetS_1\right\vert,
\\
&\left\vert\symSetS_1 \cap \symSetS_2\right\vert
=  
\left\vert\symSetS_1\right\vert  
+
\left\vert\symSetS_2\right\vert
-
\left\vert\symSetS_1 \cup \symSetS_2\right\vert.
\end{aligned}
\end{equation}
It allows to express set operation cardinalities in terms of the union cardinality. However, this approach can lead to very large estimation errors, especially if the result is small compared to the operand cardinalities \cite{Dasgupta2015}. In the worst case, the estimate could be negative without artificial restriction to nonnegative values.

Motivated by the good results we have obtained for a single \ac{HLL} sketch using the \ac{ML} method in combination with the Poisson approximation, we applied the same approach also for the estimation of set operation result sizes. Assume two given \ac{HLL} sketches with register values $\boldsymbol{\symRegValVariate}_1=(\symRegValVariate_{11},\ldots,\symRegValVariate_{1\symNumReg})$ and $\boldsymbol{\symRegValVariate}_2=(\symRegValVariate_{21},\ldots,\symRegValVariate_{2\symNumReg})$ representing sets $\symSetS_1$ and $\symSetS_2$, respectively. The goal is to find estimates for the cardinalities of the pairwise disjoint sets $\symSetA = \symSetS_1\setminus\symSetS_2$, $\symSetB = \symSetS_2\setminus\symSetS_1$, and $\symSetX = \symSetS_1\cap\symSetS_2$. The Poisson approximation allows us to assume that pairwise distinct elements are inserted into the sketches representing $\symSetS_1$ and $\symSetS_2$ at rates $\symPoissonRate_\symSetASuffix$ and $\symPoissonRate_\symSetBSuffix$, respectively. Furthermore, we assume that further unique elements are inserted into both sketches simultaneously at rate $\symPoissonRate_\symSetXSuffix$. We expect that good estimates $\symPoissonRateEstimate_\symSetASuffix$, $\symPoissonRateEstimate_\symSetBSuffix$, and $\symPoissonRateEstimate_\symSetXSuffix$ for the rates are also good estimates for the cardinalities $\vert\symSetA\vert$, $\vert\symSetB\vert$, and $\vert\symSetX\vert$. 

\subsection{Joint Log-Likelihood Function}
In order to apply the \ac{ML} method, we need to find the 
joint probability distribution of both sketches. Under the Poisson model the individual registers are independent and identically distributed. Therefore, we first derive the joint probability distribution for a single register that has value $\symRegValVariate_1$ in the first sketch that represents $\symSetS_1$ and value $\symRegValVariate_2$ in the second that represents $\symSetS_2$. 
The first one can be thought to be constructed by merging two sketches representing $\symSetA$ and $\symSetX$, respectively. Analogously, the sketch for $\symSetS_2$ could have been obtained from sketches for $\symSetB$ and $\symSetX$. Let $\symRegValVariate_\symSetASuffix$, $\symRegValVariate_\symSetBSuffix$, and $\symRegValVariate_\symSetXSuffix$ be the values of the considered register in the sketches for $\symSetA$, $\symSetB$, and $\symSetX$, respectively. The corresponding values in sketches for $\symSetS_1$ and $\symSetS_2$ are given by $\symRegValVariate_1 = \max(\symRegValVariate_\symSetASuffix, \symRegValVariate_\symSetXSuffix)$ and $\symRegValVariate_2 = \max(\symRegValVariate_\symSetBSuffix, \symRegValVariate_\symSetXSuffix)$.
Since $\symSetA$, $\symSetB$, and $\symSetX$ are disjoint and therefore $\symRegValVariate_\symSetASuffix$, $\symRegValVariate_\symSetBSuffix$, and $\symRegValVariate_\symSetXSuffix$ are  independent, the joint cumulative probability function of $\symRegValVariate_1$ and $\symRegValVariate_2$ is \begin{multline}
\label{equ:joint_cdf}
\symProbability(
\symRegValVariate_1 \leq \symRegVal_1
\wedge
\symRegValVariate_2 \leq \symRegVal_2
)
=
\\
\symProbability(
\symRegValVariate_\symSetASuffix \leq \symRegVal_1)
\,
\symProbability(
\symRegValVariate_\symSetBSuffix \leq \symRegVal_2)
\,
\symProbability(
\symRegValVariate_\symSetXSuffix \leq \min(\symRegVal_1, \symRegVal_2)
)
=
\\
\begin{cases}
0 & \symRegVal_1 < 0 \vee \symRegVal_2 < 0
\\
e^{
-
\frac{\symPoissonRate_{\symSetASuffix}}{\symNumReg 2^{\symRegVal_1}}
-
\frac{\symPoissonRate_{\symSetBSuffix}}{\symNumReg 2^{\symRegVal_2}}
-
\frac{\symPoissonRate_{\symSetXSuffix}}{\symNumReg 2^{\min(\symRegVal_1, \symRegVal_2)}}
}
& 0\leq\symRegVal_1 \leq \symRegRange \wedge 0\leq\symRegVal_2\leq\symRegRange
\\
e^{
-
\frac{\symPoissonRate_{\symSetBSuffix} + \symPoissonRate_{\symSetXSuffix}}{\symNumReg 2^{\symRegVal_2}}
}
& 0\leq\symRegVal_2 \leq \symRegRange < \symRegVal_1
\\
e^{
-
\frac{\symPoissonRate_{\symSetASuffix} + \symPoissonRate_{\symSetXSuffix}}{\symNumReg 2^{\symRegVal_1}}
}
&  0\leq\symRegVal_1 \leq \symRegRange < \symRegVal_2
\\
1
&
\symRegRange < \symRegVal_1 \wedge \symRegRange < \symRegVal_2.
\end{cases}
\end{multline}
Here we used that 
$\symRegValVariate_\symSetASuffix$, $\symRegValVariate_\symSetBSuffix$, and $\symRegValVariate_\symSetXSuffix$ follow \eqref{equ:register_value_distribution} under the Poisson model. The corresponding probability mass function is
\begin{multline*}
\symProbabilityMass(\symRegVal_1,\symRegVal_2)
=
\symProbability(
\symRegValVariate_1 \leq \symRegVal_1
\wedge
\symRegValVariate_2 \leq \symRegVal_2
)
-
\symProbability(
\symRegValVariate_1 \leq \symRegVal_1-1
\wedge
\symRegValVariate_2 \leq \symRegVal_2
)
\\
-\symProbability(
\symRegValVariate_1 \leq \symRegVal_1
\wedge
\symRegValVariate_2 \leq \symRegVal_2-1
)
+\symProbability(
\symRegValVariate_1 \leq \symRegVal_1-1
\wedge
\symRegValVariate_2 \leq \symRegVal_2-1
).
\end{multline*}
Since the values for different registers are independent under the Poisson model, the joint probability mass function for all registers in both sketches is
$\symProbabilityMass(\boldsymbol{\symRegVal}_1,\boldsymbol{\symRegVal}_2)
=
\prod_{\symIndexI = 1}^{\symNumReg}
\symProbabilityMass(\symRegVal_{1\symIndexI},\symRegVal_{2\symIndexI})$.
The \ac{ML} estimates $\symPoissonRateEstimate_\symSetASuffix$,
 $\symPoissonRateEstimate_\symSetBSuffix$, and  $\symPoissonRateEstimate_\symSetXSuffix$ can be finally obtained by maximizing the log-likelihood function given by
\begingroup
\allowdisplaybreaks
\begin{multline}
\label{equ:log_likelihood_pair}
\log \symLikelihood(
\symPoissonRate_\symSetASuffix,
\symPoissonRate_\symSetBSuffix,
\symPoissonRate_\symSetXSuffix
\vert
\boldsymbol{\symRegValVariate}_1,
\boldsymbol{\symRegValVariate}_2
)
=
\sum_{\symIndexI = 1}^{\symNumReg}
\log(\symProbabilityMass(\symRegValVariate_{1\symIndexI},\symRegValVariate_{2\symIndexI}))
=
\\
\sum_{\symRegVal=1}^{\symRegRange}
\log\left(1-e^{-\frac{\symPoissonRate_\symSetASuffix+\symPoissonRate_\symSetXSuffix}{\symNumReg 2^{\symRegVal}}}\right)
\symCountVariate^{<}_{1\symRegVal}
+
\log\left(1-e^{-\frac{\symPoissonRate_\symSetBSuffix+\symPoissonRate_\symSetXSuffix}{\symNumReg 2^{\symRegVal}}}\right)
\symCountVariate^{<}_{2\symRegVal}
\\
+
\sum_{\symRegVal=1}^{\symRegRange+1}
\log\left(1-e^{-\frac{\symPoissonRate_\symSetASuffix}{\symNumReg 2^{\min\left(\symRegVal,\symRegRange\right)}}}\right)
\symCountVariate^{>}_{1\symRegVal}
+
\log\left(1-e^{-\frac{\symPoissonRate_\symSetBSuffix}{\symNumReg 2^{\min\left(\symRegVal,\symRegRange\right)}}}\right)
\symCountVariate^{>}_{2\symRegVal}
\\
+
\sum_{\symRegVal=1}^{\symRegRange+1}
\log\left(
1
-e^{-\frac{\symPoissonRate_\symSetASuffix+\symPoissonRate_\symSetXSuffix}{\symNumReg 2^{\min\left(\symRegVal,\symRegRange\right)}}}
-
e^{-\frac{\symPoissonRate_\symSetBSuffix+\symPoissonRate_\symSetXSuffix}{\symNumReg 2^{\min\left(\symRegVal,\symRegRange\right)}}}
+
e^{-\frac{\symPoissonRate_\symSetASuffix+\symPoissonRate_\symSetBSuffix+\symPoissonRate_\symSetXSuffix}{\symNumReg 2^{\min\left(\symRegVal,\symRegRange\right)}}}
\right)
\symCountVariate^{=}_{\symRegVal}
\\
-
\frac{\symPoissonRate_\symSetASuffix}{\symNumReg}
\sum_{\symRegVal=0}^{\symRegRange}
\frac{
  \symCountVariate^{<}_{1\symRegVal}+
  \symCountVariate^{=}_{\symRegVal}+
  \symCountVariate^{>}_{1\symRegVal}
}{2^{\symRegVal}}
-
\frac{\symPoissonRate_\symSetBSuffix}{\symNumReg}
\sum_{\symRegVal=0}^{\symRegRange}
\frac{
  \symCountVariate^{<}_{2\symRegVal}+
  \symCountVariate^{=}_{\symRegVal}+
  \symCountVariate^{>}_{2\symRegVal}
}{2^{\symRegVal}}
\\
-
\frac{\symPoissonRate_\symSetXSuffix}{\symNumReg}
\sum_{\symRegVal=0}^{\symRegRange}
\frac{
  \symCountVariate^{<}_{1\symRegVal}+
  \symCountVariate^{=}_{\symRegVal}+
  \symCountVariate^{<}_{2\symRegVal}
}{2^{\symRegVal}}
\end{multline}
\endgroup
where
$\symCountVariate^{<}_{1\symRegVal}$,
$\symCountVariate^{>}_{1\symRegVal}$,
$\symCountVariate^{<}_{2\symRegVal}$,
$\symCountVariate^{>}_{2\symRegVal}$,
and $\symCountVariate^{=}_{\symRegVal}$ are defined as
\begin{equation}
\label{equ:sufficient_joint_statistic}
\begin{aligned}
\symCountVariate^{<}_{1\symRegVal}&:=\left|\lbrace\symIndexI\vert\symRegVal=\symRegValVariate_{1\symIndexI}<
\symRegValVariate_{2\symIndexI}\rbrace\right|,&
\symCountVariate^{>}_{1\symRegVal}&:=\left|\lbrace\symIndexI\vert\symRegVal=\symRegValVariate_{1\symIndexI}>
\symRegValVariate_{2\symIndexI}\rbrace\right|,
\\
\symCountVariate^{<}_{2\symRegVal}&:=\left|\lbrace\symIndexI\vert\symRegVal=\symRegValVariate_{2\symIndexI}<
\symRegValVariate_{1\symIndexI}\rbrace\right|,&
\symCountVariate^{>}_{2\symRegVal}&:=\left|\lbrace\symIndexI\vert\symRegVal=\symRegValVariate_{2\symIndexI}>
\symRegValVariate_{1\symIndexI}\rbrace\right|,
\\
\symCountVariate^{=}_{\symRegVal}&:=\left|\lbrace\symIndexI\vert\symRegVal=\symRegValVariate_{1\symIndexI}=
\symRegValVariate_{2\symIndexI}\rbrace\right|
\end{aligned}
\end{equation}
for $0\leq \symRegVal \leq \symRegRange+1$. These $5(\symRegRange+2)$ values represent a sufficient statistic for estimating $\symPoissonRate_\symSetASuffix$, $\symPoissonRate_\symSetBSuffix$, and $\symPoissonRate_\symSetXSuffix$ and greatly reduce the number of terms and also the evaluation costs of the log-likelihood function. The derived formula is the generalization of \eqref{equ:log_likelihood_single} to two sketches and therefore has a similar structure.

\subsection{Numerical Optimization}
The \ac{ML} estimates $\symPoissonRateEstimate_\symSetASuffix$, $\symPoissonRateEstimate_\symSetBSuffix$, and $\symPoissonRateEstimate_\symSetXSuffix$ are obtained by maximizing \eqref{equ:log_likelihood_pair}. Since the three parameters are all nonnegative, this is a constrained optimization problem. The transformation $\symPoissonRate=e^{\symPhi}$ helps to get rid of these constraints and also translates relative accuracy limits into absolute ones, because $\Delta\symPhi = \Delta\symPoissonRate/\symPoissonRate$. Many optimizer implementations allow the definition of absolute limits rather than relative ones. Quasi-Newton methods are commonly used to find the maximum of such multi-dimensional functions. They all require the computation of the gradient which can be straightforwardly derived for \eqref{equ:log_likelihood_pair}.  Among these methods the \ac{BFGS} algorithm \cite{Press2007} is very popular. We used the implementation provided by the Dlib C++ library \cite{King2009} for our experiments. Good initial guess values are important to ensure fast convergence for any optimization algorithm. An obvious choice are the cardinality estimates obtained by application of the inclusion-exclusion principle \eqref{equ:conventional_approach}. However, in order to ensure that their logarithms are all defined, we require that the initial values are not smaller than 1. The optimization loop is continued until the relative changes of $\symPoissonRate_\symSetASuffix$, $\symPoissonRate_\symSetBSuffix$, and $\symPoissonRate_\symSetXSuffix$ are all smaller than a predefined threshold. For the results presented below we used again $\symStopDelta=\symStopEpsilon/\sqrt{\symNumReg}$ with $\symStopEpsilon = 10^{-2}$. A few tens of iterations are typically necessary to satisfy this stop criterion for starting points determined by the inclusion-exclusion principle.

\subsection{Results}
\begin{table*}
\centering
\caption{The cardinalities of \boldmath$\symSetA=\symSetS_1\setminus\symSetS_2$, $\symSetB=\symSetS_2\setminus\symSetS_1$, $\symSetX=\symSetS_1\cap\symSetS_2$,
and $\symSetU=\symSetS_1\cup\symSetS_2$ have been estimated from 3000 different \ac{HLL} sketch pairs with $\symPrecision=16$ and $\symRegRange=16$ representing randomly generated sets $\symSetS_1$ and $\symSetS_2$ with fixed intersection and relative complement cardinalities. We have determined the \acs{RMSE} for the inclusion-exclusion principle and the \ac{ML} approach together with the corresponding improvement factor for 40 different cases.}
\label{tbl:joint_estimation_results}
\begin{filecontents*}{joint_cardinality_calculation.csv}
trueCardA,trueCardB,trueCardX,trueJaccardIdx,trueRatio,inclExclMinA,inclExclMaxA,inclExclMeanA,inclExclStdDevA,inclExclRMSEA,inclExclMinB,inclExclMaxB,inclExclMeanB,inclExclStdDevB,inclExclRMSEB,inclExclMinX,inclExclMaxX,inclExclMeanX,inclExclStdDevX,inclExclRMSEX,inclExclMinAX,inclExclMaxAX,inclExclMeanAX,inclExclStdDevAX,inclExclRMSEAX,inclExclMinBX,inclExclMaxBX,inclExclMeanBX,inclExclStdDevBX,inclExclRMSEBX,inclExclMinABX,inclExclMaxABX,inclExclMeanABX,inclExclStdDevABX,inclExclRMSEABX,inclExclMinJaccardIdx,inclExclMaxJaccardIdx,inclExclMeanJaccardIdx,inclExclStdDevJaccardIdx,inclExclRMSEJaccardIdx,maxLikeMinA,maxLikeMaxA,maxLikeMeanA,maxLikeStdDevA,maxLikeRMSEA,maxLikeMinB,maxLikeMaxB,maxLikeMeanB,maxLikeStdDevB,maxLikeRMSEB,maxLikeMinX,maxLikeMaxX,maxLikeMeanX,maxLikeStdDevX,maxLikeRMSEX,maxLikeMinAX,maxLikeMaxAX,maxLikeMeanAX,maxLikeStdDevAX,maxLikeRMSEAX,maxLikeMinBX,maxLikeMaxBX,maxLikeMeanBX,maxLikeStdDevBX,maxLikeRMSEBX,maxLikeMinABX,maxLikeMaxABX,maxLikeMeanABX,maxLikeStdDevABX,maxLikeRMSEABX,maxLikeMinJaccardIdx,maxLikeMaxJaccardIdx,maxLikeMeanJaccardIdx,maxLikeStdDevJaccardIdx,maxLikeRMSEJaccardIdx,improvementStdevA,improvementRmseA,improvementStdevB,improvementRmseB,improvementStdevX,improvementRmseX,improvementStdevAX,improvementRmseAX,improvementStdevBX,improvementRmseBX,improvementStdevABX,improvementRmseABX,improvementStdevJaccard,improvementRmseJaccard,avgNumFunctionEvaluations,avgNumGradientEvaluations
69051,43258,818,0.00723081,1.59626,-0.0165356,0.0214296,1.77498e-06,0.00483224,0.00483224,-0.0225474,0.0294445,9.53492e-05,0.00677409,0.00677476,-1,1.17034,-0.00292365,0.318572,0.318585,-0.0118593,0.0105485,-3.24748e-05,0.00295434,0.00295452,-0.0108594,0.0103996,3.93201e-05,0.00288378,0.00288405,-0.00997635,0.0141987,1.64031e-05,0.00315912,0.00315916,-1,1.18976,-0.00218217,0.320907,0.320915,-0.0158453,0.0126426,-4.43045e-05,0.00334723,0.00334752,-0.0133449,0.012368,2.18692e-05,0.00380454,0.0038046,-0.462414,0.464371,0.000795148,0.129963,0.129965,-0.0118565,0.01057,-3.44765e-05,0.00295227,0.00295247,-0.0108321,0.010434,3.62204e-05,0.00287863,0.00287885,-0.00901786,0.00841324,-1.29308e-05,0.00229688,0.00229692,-0.462939,0.473731,0.000931507,0.130912,0.130915,1.44365,1.44353,1.78053,1.78068,2.45125,2.45131,1.0007,1.00069,1.00179,1.0018,1.37539,1.37539,2.45132,2.45132,25.3497,25.3497
429886036,170398425,45365204,0.0702629,2.52283,-0.022729,0.0224722,0.000223203,0.00575772,0.00576204,-0.0322148,0.0374561,0.000534793,0.0104205,0.0104342,-0.128188,0.125712,-0.00159528,0.0358452,0.0358807,-0.013956,0.0136199,4.96197e-05,0.00406124,0.00406155,-0.0149407,0.0138687,8.69369e-05,0.00399613,0.00399707,-0.0149673,0.0150389,0.000177666,0.00409365,0.00409751,-0.13497,0.132338,-0.00167384,0.0382659,0.0383025,-0.0157452,0.0155541,5.9989e-05,0.00459711,0.0045975,-0.0310924,0.0243863,0.000147214,0.00662343,0.00662506,-0.0666398,0.0663201,-0.000163423,0.0204535,0.0204542,-0.0137524,0.0136534,3.86631e-05,0.00403891,0.0040391,-0.0149536,0.0138506,8.19011e-05,0.00399139,0.00399223,-0.0114901,0.0116736,6.73116e-05,0.00334595,0.00334663,-0.0711551,0.0695844,-0.000206676,0.0213341,0.0213351,1.25246,1.2533,1.57327,1.57496,1.75252,1.7542,1.00553,1.00556,1.00119,1.00121,1.22347,1.22437,1.79365,1.79528,30.0037,30.0037
3808040,680932,1530,0.000340719,5.59239,-0.0185053,0.0171988,-0.000110791,0.00470497,0.00470628,-0.0422437,0.0480496,-7.57927e-05,0.0146048,0.014605,-1,19.8907,2.0376,3.99827,4.48753,-0.0142371,0.0167902,0.000707595,0.00425356,0.00431202,-0.014694,0.0457,0.00449244,0.00841585,0.00953984,-0.0164976,0.018769,0.000588803,0.00463579,0.00467304,-1,19.9453,2.04492,4.01193,4.50303,-0.0146703,0.0129068,-6.14412e-05,0.00408415,0.00408462,-0.0183943,0.0171203,0.000192602,0.00489618,0.00489997,-1,5.76169,-0.082873,1.29665,1.29929,-0.0142391,0.0126462,-9.47e-05,0.00405743,0.00405853,-0.0146934,0.0148415,6.37835e-06,0.00387815,0.00387815,-0.0131262,0.0115314,-5.1134e-05,0.00352298,0.00352335,-1,5.73348,-0.0822991,1.29785,1.30046,1.15201,1.1522,2.9829,2.98064,3.08354,3.45383,1.04834,1.06246,2.17007,2.45989,1.31587,1.32631,3.09121,3.46265,31.429,31.429
397877,18569,48262,0.103854,21.4269,-0.0150882,0.0152124,-7.72961e-05,0.00436083,0.00436152,-0.107062,0.128716,-0.000490695,0.0274773,0.0274816,-0.0558463,0.0374387,0.000257335,0.0108298,0.0108328,-0.0121093,0.0125485,-4.10967e-05,0.00373934,0.00373957,-0.0116196,0.0113014,4.94947e-05,0.0030261,0.0030265,-0.0125715,0.0137858,-5.9062e-05,0.00375814,0.0037586,-0.0659556,0.0447768,0.000340504,0.012302,0.0123067,-0.0137939,0.0138036,-6.57779e-06,0.00405664,0.00405665,-0.0964027,0.0720975,0.000256436,0.0199178,0.0199195,-0.0330074,0.0326763,-2.66915e-05,0.00813176,0.0081318,-0.0128951,0.0132746,-8.75364e-06,0.00364065,0.00364066,-0.0116057,0.0113445,5.19754e-05,0.00302359,0.00302403,-0.0118327,0.0125294,1.84291e-06,0.00349747,0.00349747,-0.0338019,0.0326768,-1.4898e-05,0.00900856,0.00900857,1.07499,1.07515,1.37953,1.37964,1.33179,1.33215,1.02711,1.02717,1.00083,1.00082,1.07453,1.07466,1.36559,1.36611,28.5103,28.5103
239529,24778,326,0.00123189,9.667,-0.0122772,0.0161402,8.59281e-05,0.00396745,0.00396838,-0.0656532,0.0586171,0.000272212,0.016855,0.0168572,-1,4.77507,0.144227,1.02296,1.03308,-0.0117228,0.0147591,0.000281838,0.00358896,0.00360001,-0.008925,0.0448699,0.0021416,0.00622798,0.00658591,-0.0112299,0.0188656,0.000280937,0.00379903,0.0038094,-1,4.80333,0.145729,1.0255,1.0358,-0.0118291,0.0129232,0.000101478,0.0035995,0.00360093,-0.0220905,0.0218645,0.000426959,0.00657199,0.00658585,-1,1.38499,-0.0326396,0.445235,0.446429,-0.0116878,0.0118545,5.69775e-05,0.00352941,0.00352987,-0.0100929,0.00924135,-2.44323e-06,0.00286104,0.00286104,-0.0106789,0.0119574,9.16194e-05,0.00327167,0.00327295,-1,1.38035,-0.0324393,0.445775,0.446954,1.10222,1.10204,2.56468,2.55961,2.29758,2.31409,1.01687,1.01987,2.17682,2.30193,1.16119,1.1639,2.30049,2.31747,32.2577,32.2577
165754,53843,108,0.000491568,3.07847,-0.0150162,0.0169982,0.000138472,0.00456951,0.00457161,-0.0356502,0.0327089,0.000357672,0.00981825,0.00982477,-1,16.031,1.29851,2.99156,3.26122,-0.0124917,0.016336,0.000983896,0.00373576,0.00386316,-0.00927741,0.0306416,0.00295633,0.00573021,0.00644788,-0.0121997,0.0173996,0.000830428,0.00429084,0.00437046,-1,16.2086,1.30451,3.00325,3.27433,-0.0122516,0.0131701,-5.13963e-05,0.00343143,0.00343182,-0.0136406,0.0125138,-0.000227058,0.00368652,0.0036935,-1,4.54643,0.101719,1.09376,1.09848,-0.0124909,0.0125104,1.48711e-05,0.00334956,0.0033496,-0.0100075,0.0105023,-2.29804e-05,0.00293217,0.00293226,-0.0100529,0.0103828,-4.44185e-05,0.00266614,0.00266651,-1,4.56265,0.102392,1.09511,1.09988,1.33166,1.33212,2.66329,2.66001,2.73511,2.96884,1.1153,1.15332,1.95426,2.19895,1.60938,1.63902,2.74243,2.97698,38.341,38.341
1667351,3974,142771,0.0787009,419.565,-0.0150904,0.0151009,-5.78515e-05,0.00436182,0.0043622,-0.410865,0.471005,0.00242289,0.121111,0.121135,-0.0156056,0.0145469,-0.00014822,0.00473882,0.00474114,-0.0139547,0.0141536,-6.49792e-05,0.00401454,0.00401506,-0.0116037,0.0116573,-7.85916e-05,0.00331303,0.00331397,-0.013795,0.0138272,-5.95292e-05,0.00401652,0.00401697,-0.021264,0.0193915,-7.24127e-05,0.00623621,0.00623663,-0.0149378,0.0146291,-6.37325e-05,0.00422256,0.00422304,-0.314598,0.391419,0.00417931,0.0976084,0.0976978,-0.015612,0.0163426,-0.000196771,0.00429204,0.00429655,-0.0137769,0.0131752,-7.42258e-05,0.00389865,0.00389935,-0.0116008,0.0116506,-7.82626e-05,0.00331305,0.00331397,-0.0136358,0.0131524,-6.49079e-05,0.00388997,0.00389051,-0.0212348,0.01998,-0.000117358,0.00568503,0.00568624,1.03298,1.03295,1.24078,1.23989,1.1041,1.10348,1.02973,1.02967,0.999996,0.999998,1.03253,1.0325,1.09695,1.09679,27.6163,27.6163
69742,1058,115,0.00162166,65.9187,-0.0112818,0.0104188,-6.02936e-05,0.00302774,0.00302834,-0.150627,0.123465,-0.000593943,0.0368461,0.0368509,-1,1.35416,0.00584742,0.336489,0.33654,-0.0114326,0.00992968,-5.05682e-05,0.00296281,0.00296324,-0.0108318,0.0133217,3.75637e-05,0.00275388,0.00275413,-0.0110015,0.0102257,-5.8675e-05,0.0029784,0.00297898,-1,1.36067,0.00611534,0.33714,0.337196,-0.0112268,0.0105937,-5.13809e-05,0.00297957,0.00298001,-0.0610904,0.0587091,-2.31974e-05,0.0188973,0.0188973,-0.552245,0.556032,0.000501357,0.171452,0.171453,-0.0114302,0.00993636,-5.0471e-05,0.00296169,0.00296212,-0.0108173,0.0082742,2.82296e-05,0.00273876,0.00273891,-0.0109474,0.0103977,-5.00641e-05,0.00293061,0.00293104,-0.553239,0.553883,0.000606952,0.171757,0.171758,1.01617,1.01622,1.94981,1.95006,1.96258,1.96287,1.00038,1.00038,1.00552,1.00556,1.01631,1.01636,1.96289,1.9632,21.1833,21.1833
871306344,3808040,66873869,0.0709923,228.807,-0.0169462,0.0140634,0.000136546,0.00442645,0.00442856,-0.30252,0.299583,-0.000438971,0.0890235,0.0890245,-0.0269785,0.0230307,9.85833e-06,0.00651068,0.00651069,-0.0151899,0.0132694,0.000127516,0.00409529,0.00409728,-0.0155395,0.014759,-1.43227e-05,0.00405018,0.0040502,-0.0151821,0.0131431,0.000125226,0.00410464,0.00410655,-0.030442,0.0299858,-9.73499e-05,0.00784478,0.00784538,-0.0164432,0.0131862,0.000114348,0.00427121,0.00427274,-0.233713,0.24721,-0.00100363,0.0727645,0.0727714,-0.0227187,0.0235066,4.20345e-05,0.00582529,0.00582544,-0.0148559,0.0122017,0.000109194,0.00397285,0.00397435,-0.0155344,0.0147662,-1.43013e-05,0.0040501,0.00405013,-0.0147164,0.0120611,0.000104695,0.00396082,0.0039622,-0.0234541,0.0269789,-4.7346e-05,0.00699003,0.00699019,1.03635,1.03647,1.22345,1.22334,1.11766,1.11763,1.03082,1.03093,1.00002,1.00002,1.03631,1.03643,1.12228,1.12234,27.6773,27.6773
362986091,32023944,6781734,0.0168787,11.3348,-0.0177504,0.0154805,0.000192261,0.00448451,0.00448863,-0.0598218,0.0659923,0.000966421,0.0197582,0.0197818,-0.325776,0.281612,-0.00406372,0.0919311,0.0920208,-0.0142007,0.0148589,0.000114205,0.00408511,0.0040867,-0.0140396,0.0138531,8.73466e-05,0.0040736,0.00407454,-0.0161045,0.014264,0.000182129,0.00406486,0.00406894,-0.331326,0.29023,-0.00409225,0.0934593,0.0935489,-0.0151271,0.0151778,0.000140386,0.00418099,0.00418335,-0.0387356,0.0429743,0.000428804,0.0107518,0.0107604,-0.199724,0.168522,-0.00152691,0.0471966,0.0472213,-0.0140985,0.0146824,0.000109807,0.00408027,0.00408175,-0.0140364,0.0138649,8.70199e-05,0.00407413,0.00407506,-0.013731,0.0139919,0.000135232,0.00379544,0.00379785,-0.200084,0.175844,-0.00162543,0.0478121,0.0478397,1.0726,1.07298,1.83766,1.83839,1.94783,1.94872,1.00118,1.00121,0.99987,0.999872,1.07098,1.07138,1.95472,1.95546,29.052,29.052
9050248,342712,71854,0.0075917,26.4077,-0.0134412,0.0140821,3.69887e-06,0.00415485,0.00415486,-0.110059,0.120098,6.22114e-05,0.0304426,0.0304426,-0.551815,0.497864,-0.000479227,0.144132,0.144133,-0.0136069,0.0136663,-1.05096e-07,0.00399972,0.00399972,-0.0114895,0.0117542,-3.16325e-05,0.00368238,0.00368252,-0.0128069,0.0138038,2.15133e-06,0.00397764,0.00397764,-0.554606,0.505289,-0.000326441,0.145112,0.145112,-0.0139607,0.0134979,1.58377e-06,0.00403141,0.00403141,-0.0539079,0.0561577,-2.24573e-05,0.014835,0.014835,-0.261523,0.269297,-7.51316e-05,0.0686512,0.0686513,-0.0136262,0.0136722,9.79487e-07,0.00400133,0.00400133,-0.0114824,0.0117613,-3.1587e-05,0.00368246,0.00368259,-0.0132366,0.0131112,1.30863e-07,0.00385957,0.00385957,-0.266493,0.276011,-4.3886e-05,0.0690041,0.0690041,1.03062,1.03062,2.05208,2.05209,2.09949,2.0995,0.999598,0.999598,0.999979,0.999979,1.03059,1.03059,2.10295,2.10295,28.616,28.616
708580,1038,42830,0.0569209,682.64,-0.0139724,0.0164921,-3.00736e-05,0.00412937,0.00412948,-0.456301,0.572966,-3.31759e-05,0.147838,0.147838,-0.0167821,0.0148719,-3.15929e-05,0.00464534,0.00464545,-0.0133858,0.015684,-3.01602e-05,0.00389205,0.00389217,-0.00895994,0.0100706,-3.16304e-05,0.00291106,0.00291123,-0.0132733,0.0154849,-3.01644e-05,0.00389199,0.00389211,-0.023007,0.0218911,1.39613e-05,0.00610008,0.00610009,-0.0137964,0.0150031,-1.06274e-05,0.00403653,0.00403654,-0.36518,0.41398,0.00240072,0.116755,0.11678,-0.0189275,0.0154136,-9.02944e-05,0.00408453,0.00408553,-0.0132151,0.0142495,-1.51684e-05,0.00381242,0.00381245,-0.0089559,0.010063,-3.13523e-05,0.00291113,0.0029113,-0.0131077,0.0140831,-1.18356e-05,0.00380536,0.00380538,-0.0190047,0.0209657,-6.44078e-05,0.00550529,0.00550567,1.023,1.02302,1.26622,1.26595,1.1373,1.13705,1.02089,1.02091,0.999976,0.999977,1.02276,1.02279,1.10804,1.10797,24.711,24.711
50111408,54381,38388,0.000764638,921.487,-0.0154726,0.0145929,-0.000113945,0.0041128,0.00411437,-0.612799,0.628744,0.000838038,0.172685,0.172687,-0.896883,0.861357,-0.00128972,0.244596,0.2446,-0.0152333,0.014354,-0.000114845,0.00410958,0.00411118,-0.0101933,0.0125103,-4.24343e-05,0.00308977,0.00309006,-0.0154395,0.0145652,-0.000113813,0.00410528,0.00410685,-0.896131,0.851165,-0.00113593,0.244796,0.244799,-0.0153115,0.0145677,-0.000112472,0.00411455,0.00411609,-0.483988,0.457257,0.000984587,0.107427,0.107432,-0.641463,0.684393,-0.0014974,0.152235,0.152242,-0.0151165,0.014385,-0.000113532,0.00410969,0.00411126,-0.0101933,0.0125106,-4.24635e-05,0.00308978,0.00309007,-0.0152787,0.01454,-0.000112342,0.00410703,0.00410856,-0.639845,0.697551,-0.00135025,0.152454,0.15246,0.999573,0.999583,1.60746,1.60741,1.6067,1.60665,0.999973,0.999982,0.999996,0.999996,0.999574,0.999584,1.60571,1.60566,31.3907,31.3907
647883,574964,8380,0.00680622,1.12682,-0.0226474,0.0260174,3.26151e-05,0.00664323,0.00664331,-0.0215653,0.0244273,3.26674e-06,0.0070899,0.0070899,-1,1.41478,-0.00174386,0.408169,0.408172,-0.0127979,0.0140863,9.93081e-06,0.00386599,0.00386601,-0.012737,0.0125901,-2.18316e-05,0.00378983,0.00378989,-0.0126636,0.0178649,6.81879e-06,0.00393277,0.00393278,-1,1.43883,-0.000581347,0.411004,0.411004,-0.0138321,0.0142974,4.88999e-05,0.00428814,0.00428842,-0.0170116,0.0164759,2.05753e-05,0.00434595,0.00434599,-0.999881,0.490695,-0.00409958,0.146895,0.146952,-0.0127942,0.0140219,-4.07319e-06,0.00385927,0.00385927,-0.0127506,0.0125568,-3.86126e-05,0.00378076,0.00378095,-0.00856247,0.0121066,7.43729e-06,0.00286062,0.00286063,-0.999881,0.494363,-0.00396889,0.147737,0.14779,1.54921,1.54913,1.63138,1.63136,2.77864,2.77759,1.00174,1.00175,1.0024,1.00236,1.3748,1.37479,2.782,2.781,29.1317,29.1317
535085964,55907711,49124015,0.0767422,9.57088,-0.0170464,0.0217229,9.19989e-05,0.0047969,0.00479778,-0.0647796,0.0640252,0.000113977,0.0190866,0.0190869,-0.0728919,0.0721366,9.05781e-05,0.0214555,0.0214557,-0.0154148,0.0167464,9.18794e-05,0.00407127,0.00407231,-0.012854,0.0138039,0.000103033,0.00408719,0.00408848,-0.0150349,0.0184067,9.38094e-05,0.00409585,0.00409693,-0.0768727,0.0784726,4.35091e-05,0.0231684,0.0231685,-0.0155053,0.0186088,8.55494e-05,0.0043701,0.00437094,-0.0448155,0.0421273,0.000125325,0.012845,0.0128456,-0.0448728,0.0509401,8.00903e-05,0.0142203,0.0142205,-0.0142542,0.0163758,8.50904e-05,0.00401621,0.00401711,-0.0129129,0.0138543,0.000104168,0.00408634,0.00408767,-0.0123938,0.0157819,8.86044e-05,0.00375002,0.00375107,-0.0513927,0.0533754,1.02317e-05,0.0150212,0.0150212,1.09766,1.09765,1.48592,1.48587,1.5088,1.50879,1.01371,1.01374,1.00021,1.0002,1.09222,1.0922,1.54239,1.54239,29.3233,29.3233
6125762237,23524057,142456034,0.0226417,260.404,-0.0162081,0.0155158,7.41628e-06,0.0043639,0.0043639,-0.320801,0.444274,-0.00267168,0.107614,0.107647,-0.0729883,0.0549399,0.000340184,0.0182876,0.0182907,-0.0155653,0.0152497,1.4979e-05,0.00424825,0.00424828,-0.0146224,0.0135079,-8.66815e-05,0.00399609,0.00399703,-0.0158135,0.015185,4.93392e-06,0.00425063,0.00425063,-0.0736973,0.0568757,0.000359723,0.0191117,0.0191151,-0.0165657,0.0143999,-6.10747e-06,0.00433245,0.00433245,-0.316776,0.340452,-0.00245199,0.0978225,0.0978532,-0.0574376,0.0585875,0.000304061,0.0166859,0.0166887,-0.0157863,0.0144084,9.41636e-07,0.00422417,0.00422417,-0.0146254,0.0135077,-8.65491e-05,0.00399617,0.0039971,-0.0161617,0.0141279,-8.22957e-06,0.00422022,0.00422023,-0.0606905,0.0599122,0.000334635,0.0174718,0.017475,1.00726,1.00726,1.1001,1.10009,1.09599,1.096,1.0057,1.00571,0.999981,0.999982,1.0072,1.0072,1.09386,1.09385,27.1407,27.1407
13474584,80968,7363,0.000542877,166.419,-0.0178269,0.0128716,-6.76028e-05,0.00408331,0.00408387,-0.222038,0.254051,-0.00224989,0.0743881,0.0744221,-1,2.42673,0.0694066,0.735688,0.738954,-0.0179059,0.0132446,-2.96603e-05,0.00406044,0.00406055,-0.0113218,0.149517,0.00372317,0.0150365,0.0154906,-0.0177268,0.0127885,-4.29147e-05,0.00406527,0.0040655,-1,2.41222,0.0697961,0.736074,0.739376,-0.0179375,0.0133307,-2.94043e-05,0.00406971,0.00406982,-0.0973229,0.0993221,0.00395423,0.0385329,0.0387352,-1,1.10843,-0.0436412,0.422934,0.42518,-0.0179134,0.013263,-5.32223e-05,0.00405808,0.00405843,-0.0113221,0.00973261,-1.31768e-05,0.00307679,0.00307682,-0.0178366,0.0132446,-2.92986e-05,0.00404327,0.00404338,-1,1.10298,-0.0434795,0.423117,0.425345,1.00334,1.00345,1.93051,1.9213,1.73948,1.73798,1.00058,1.00052,4.88706,5.0346,1.00544,1.00547,1.73965,1.7383,30.5583,30.5583
1300149,56589,15219,0.0110929,22.9753,-0.0157315,0.0187581,-9.45698e-05,0.00424879,0.00424985,-0.0909852,0.0958597,0.00041635,0.0273456,0.0273487,-0.36068,0.332904,-0.00151498,0.101255,0.101267,-0.0146419,0.0169395,-0.000111004,0.0040213,0.00402283,-0.00969473,0.0101574,7.024e-06,0.00299709,0.0029971,-0.0147678,0.0178109,-8.92524e-05,0.00402404,0.00402503,-0.363488,0.339841,-0.00129266,0.102486,0.102494,-0.0151142,0.0177001,-0.000104125,0.00408579,0.00408711,-0.0498931,0.0541415,0.000233318,0.0145163,0.0145181,-0.191112,0.198637,-0.000834717,0.0528712,0.0528777,-0.0148356,0.0167588,-0.000112578,0.0040185,0.00402008,-0.00969589,0.0101551,6.95836e-06,0.00299689,0.0029969,-0.0143732,0.0168081,-9.83113e-05,0.0038702,0.00387145,-0.197056,0.196554,-0.0006998,0.0534243,0.0534288,1.0399,1.03982,1.88379,1.88376,1.91514,1.91511,1.0007,1.00068,1.00007,1.00007,1.03975,1.03967,1.91834,1.91833,28.1103,28.1103
291621648,115593125,141357,0.000347011,2.52283,-0.0219436,0.0182005,-4.22971e-05,0.00544399,0.00544415,-0.0330337,0.0356807,-7.49812e-06,0.00978792,0.00978792,-1,23.5957,2.46104,4.57759,5.19722,-0.0177672,0.0177072,0.00115008,0.00453548,0.00467902,-0.0149183,0.0344157,0.0029984,0.00637287,0.007043,-0.0158769,0.0182217,0.0008216,0.00495118,0.00501888,-1,23.7763,2.47134,4.59758,5.2197,-0.0188812,0.0122793,-0.000101481,0.00416805,0.00416928,-0.0188795,0.0142519,-0.00015961,0.00463245,0.0046352,-1,9.86561,0.146959,1.77197,1.77806,-0.0177626,0.012772,-3.02315e-05,0.00405386,0.00405397,-0.0149183,0.0134614,2.00796e-05,0.00411162,0.00411166,-0.01316,0.00986702,-6.69444e-05,0.00321748,0.00321818,-1,9.95999,0.148255,1.77525,1.78143,1.30612,1.30578,2.1129,2.11165,2.58333,2.92298,1.1188,1.15418,1.54997,1.71293,1.53884,1.55954,2.58982,2.93007,29.2133,29.2133
730051,39948,147097,0.160394,18.275,-0.017131,0.0165767,2.16183e-05,0.00486925,0.0048693,-0.0776526,0.0919285,9.03185e-05,0.0263905,0.0263906,-0.0281431,0.0243652,-0.000101423,0.00788666,0.00788731,-0.0138463,0.0148623,9.84338e-07,0.00390883,0.00390883,-0.0110367,0.0121704,-6.0472e-05,0.00333067,0.00333122,-0.0141705,0.014207,4.87567e-06,0.00393984,0.00393984,-0.0332047,0.0282559,-8.48274e-05,0.00946661,0.00946699,-0.0155839,0.0154008,2.24115e-05,0.00439632,0.00439638,-0.0686586,0.0798354,0.000101973,0.0199155,0.0199157,-0.0218697,0.0213594,-0.000104395,0.00634673,0.00634759,-0.0126117,0.0138865,1.14618e-06,0.00372363,0.00372363,-0.0110606,0.0121902,-6.03202e-05,0.00333212,0.00333266,-0.0129404,0.0130757,5.53811e-06,0.0035691,0.0035691,-0.0245143,0.0224208,-9.76466e-05,0.00721856,0.00721922,1.10757,1.10757,1.32512,1.32511,1.24263,1.24257,1.04974,1.04974,0.999567,0.999568,1.10388,1.10388,1.31143,1.31136,29.1503,29.1503
8193072,232484,13915,0.0016488,35.2414,-0.0141269,0.0134168,2.9436e-06,0.00421259,0.00421259,-0.114199,0.11442,-0.000284458,0.0334291,0.0334304,-1,1.91373,0.0128801,0.539002,0.539155,-0.0137392,0.0135213,2.4777e-05,0.00410146,0.00410154,-0.0128511,0.0514848,0.000458991,0.00449471,0.00451809,-0.0137632,0.0131492,1.62584e-05,0.0040978,0.00409783,-1,1.91447,0.0133804,0.540001,0.540167,-0.013725,0.0132698,3.8874e-05,0.0041194,0.00411958,-0.0553421,0.0656922,0.000936695,0.0152275,0.0152563,-0.999928,0.865552,-0.0148404,0.248807,0.249249,-0.0137292,0.0135426,1.36461e-05,0.00410198,0.004102,-0.0128482,0.0125854,4.57085e-05,0.0035253,0.0035256,-0.0133995,0.0128609,3.90736e-05,0.00400137,0.00400156,-0.999929,0.86656,-0.0147885,0.249053,0.249492,1.02262,1.02258,2.19531,2.19125,2.16635,2.16312,0.999874,0.999886,1.27499,1.28151,1.0241,1.02406,2.16821,2.16507,29.7833,29.7833
36810697,730051,170777,0.00452851,50.4221,-0.0143625,0.013796,-0.000105134,0.00414687,0.0041482,-0.120832,0.139162,0.000831093,0.0406249,0.0406334,-0.56968,0.523039,-0.00360796,0.173486,0.173523,-0.0150521,0.0129608,-0.00012131,0.00406013,0.00406194,-0.0150407,0.0138711,-1.04525e-05,0.00394614,0.00394616,-0.0141095,0.0134868,-0.000102872,0.00404857,0.00404987,-0.569336,0.522163,-0.00336298,0.174293,0.174325,-0.0140301,0.0138857,-0.000128381,0.00410037,0.00410238,-0.0701277,0.0621615,-0.000238867,0.020983,0.0209844,-0.268162,0.299922,0.000966193,0.0889524,0.0889576,-0.0150218,0.0130411,-0.000123326,0.00406363,0.0040655,-0.0150379,0.0138709,-1.04147e-05,0.00394614,0.00394616,-0.0137849,0.0135743,-0.000125563,0.00400339,0.00400536,-0.270185,0.303378,0.00114058,0.0894289,0.0894361,1.01134,1.01117,1.93608,1.93636,1.95032,1.95063,0.99914,0.999126,1,1,1.01128,1.01111,1.94896,1.94916,28.7097,28.7097
504075182,421415583,23291146,0.0245485,1.19615,-0.0204916,0.0246978,-4.26114e-05,0.00681399,0.00681413,-0.0264591,0.0255699,4.83175e-05,0.00744632,0.00744648,-0.420685,0.421862,6.00321e-05,0.115438,0.115438,-0.0138828,0.0167976,-3.80782e-05,0.00409496,0.00409514,-0.0139932,0.016189,4.8931e-05,0.00406099,0.00406128,-0.0143416,0.0150663,2.95741e-07,0.00403695,0.00403695,-0.427576,0.428536,0.000386334,0.118152,0.118153,-0.0165472,0.0234676,-7.11893e-05,0.00495117,0.00495168,-0.0161429,0.019872,1.20542e-05,0.00507878,0.00507879,-0.198123,0.233884,0.000698635,0.0565383,0.0565427,-0.0140623,0.016774,-3.719e-05,0.00409426,0.00409443,-0.0141294,0.0162135,4.80133e-05,0.00406197,0.00406225,-0.0117717,0.0130622,-1.53175e-05,0.00322084,0.00322088,-0.203831,0.236907,0.000789587,0.0577715,0.0577769,1.37624,1.37612,1.46616,1.46619,2.04176,2.0416,1.00017,1.00017,0.99976,0.999762,1.25338,1.25337,2.04516,2.04498,28.57,28.57
9898122,11180,6343,0.000639696,885.342,-0.0148742,0.0143322,-1.90738e-05,0.00400883,0.00400888,-0.529412,0.693933,0.00412012,0.170137,0.170187,-1,0.936358,-0.00726128,0.299197,0.299285,-0.0149522,0.0144862,-2.37118e-05,0.00400298,0.00400305,-0.00843037,0.0807609,2.64517e-07,0.00348287,0.00348287,-0.0148429,0.0143041,-1.90396e-05,0.00400156,0.00400161,-1,0.935175,-0.00717711,0.299381,0.299467,-0.0150181,0.0145115,-2.63035e-05,0.00400596,0.00400604,-0.39793,0.358199,-0.00123728,0.101239,0.101247,-0.626395,0.712498,0.00198924,0.178333,0.178344,-0.0150087,0.0145433,-2.50127e-05,0.00400347,0.00400355,-0.00843033,0.0100727,-6.93401e-05,0.00278088,0.00278175,-0.0149866,0.0144831,-2.63796e-05,0.00399873,0.00399881,-0.627035,0.70926,0.0020414,0.178445,0.178456,1.00072,1.00071,1.68055,1.68091,1.67774,1.67813,0.999879,0.999877,1.25243,1.25204,1.00071,1.0007,1.67772,1.6781,27.6757,27.6757
2287410615,484406343,4290994,0.00154569,4.72209,-0.0158445,0.0176191,3.23759e-05,0.00489212,0.00489223,-0.0421672,0.0468922,-0.000282268,0.0133144,0.0133174,-1,4.41328,0.232461,1.14703,1.17035,-0.013564,0.0157137,0.000467576,0.0042005,0.00422644,-0.0144521,0.0377,0.00176132,0.00597795,0.00623203,-0.0128465,0.0187509,0.000336735,0.00445753,0.00447023,-1,4.44663,0.234823,1.15114,1.17485,-0.0142148,0.0146016,0.000326798,0.00438546,0.00439762,-0.028323,0.0200849,0.00109989,0.00842307,0.00849458,-1,2.68251,-0.123965,0.833348,0.842518,-0.0135928,0.0145271,9.40727e-05,0.00408719,0.00408827,-0.0144219,0.0164921,1.75943e-06,0.00413959,0.00413959,-0.0119082,0.0125521,0.000269579,0.00370253,0.00371233,-1,2.69744,-0.123121,0.835107,0.844134,1.11553,1.11247,1.58071,1.56776,1.37641,1.38911,1.02772,1.0338,1.44409,1.50547,1.20391,1.20416,1.37843,1.39178,23.9123,23.9123
193929303,21724073,98798,0.000457924,8.92693,-0.0145818,0.0166875,0.000108453,0.00455765,0.00455894,-0.0582266,0.0855075,-0.000121277,0.0177038,0.0177042,-1,13.5901,1.08888,2.51359,2.73931,-0.0137057,0.0161698,0.000662851,0.00423652,0.00428806,-0.0119381,0.0805932,0.00480894,0.00945756,0.01061,-0.0135128,0.0231514,0.000583897,0.00449245,0.00453024,-1,13.6287,1.09265,2.52082,2.74744,-0.0145889,0.0149131,0.000231211,0.00414144,0.00414789,-0.0278075,0.0176267,0.000956817,0.00628085,0.00635331,-1,4.83613,-0.199269,1.0443,1.06314,-0.0137086,0.0145005,0.000129627,0.00411406,0.0041161,-0.0137034,0.0130196,5.03409e-05,0.0040003,0.00400061,-0.0135191,0.01315,0.000212917,0.003741,0.00374705,-1,4.81564,-0.198986,1.04506,1.06384,1.1005,1.0991,2.81869,2.78661,2.40697,2.57662,1.02977,1.04178,2.36422,2.65208,1.20087,1.20901,2.41213,2.58258,27.6877,27.6877
34407,4304,464,0.0118443,7.99419,-0.0120582,0.011358,-7.80227e-06,0.00321737,0.00321738,-0.0365009,0.0459273,0.000303596,0.0122581,0.0122618,-0.418288,0.359304,-0.00271614,0.110448,0.110482,-0.00946445,0.00998064,-4.38399e-05,0.00283869,0.00283903,-0.0104427,0.00868714,9.72939e-06,0.00276472,0.00276474,-0.0111379,0.0104207,-5.66849e-06,0.00284027,0.00284027,-0.421092,0.366225,-0.00256312,0.111722,0.111751,-0.0104728,0.00992376,-3.54359e-05,0.00296698,0.00296719,-0.0258566,0.0259486,0.000100496,0.00706713,0.00706785,-0.20166,0.19697,-0.000835917,0.0604795,0.0604853,-0.00944034,0.00995584,-4.60872e-05,0.00283787,0.00283825,-0.0103735,0.00870833,9.36871e-06,0.00276359,0.0027636,-0.0094279,0.00915669,-2.99827e-05,0.00261944,0.00261961,-0.203117,0.195065,-0.000761129,0.0611604,0.0611651,1.08439,1.08432,1.73452,1.73487,1.82621,1.82659,1.00029,1.00028,1.00041,1.00041,1.08431,1.08424,1.8267,1.82704,22.6983,22.6983
6714588,32737,20511,0.00303066,205.107,-0.0139401,0.0135055,2.97671e-05,0.0039832,0.00398331,-0.26431,0.352575,0.00258473,0.082133,0.0821737,-0.551378,0.424767,-0.00421185,0.131051,0.131119,-0.0138638,0.01312,1.68498e-05,0.00395612,0.00395615,-0.00893051,0.0108392,-3.32987e-05,0.00295176,0.00295195,-0.0138495,0.0133939,2.92709e-05,0.00395225,0.00395236,-0.552127,0.429316,-0.00418011,0.131449,0.131515,-0.0140258,0.013424,2.04588e-05,0.00396839,0.00396844,-0.221519,0.164821,0.00104727,0.048671,0.0486822,-0.263365,0.335188,-0.00175795,0.0776947,0.0777146,-0.0139706,0.0131499,1.50428e-05,0.00395359,0.00395362,-0.00892834,0.0108386,-3.32959e-05,0.00295177,0.00295196,-0.0139345,0.0133131,2.00358e-05,0.00393755,0.0039376,-0.260768,0.346601,-0.00174998,0.0779425,0.0779621,1.00373,1.00375,1.68752,1.68796,1.68675,1.68719,1.00064,1.00064,0.999996,0.999996,1.00374,1.00375,1.68648,1.68691,28.136,28.136
640041725,18343319,38305359,0.0549819,34.8924,-0.0146751,0.0170762,8.36656e-05,0.00436825,0.00436905,-0.116218,0.118156,-0.000647736,0.0344509,0.034457,-0.0528419,0.0588613,0.000166691,0.016652,0.0166528,-0.0132744,0.0169882,8.83539e-05,0.00405994,0.0040609,-0.0138168,0.0129798,-9.70273e-05,0.0040009,0.00400208,-0.013605,0.0158063,6.89732e-05,0.00404574,0.00404633,-0.0596601,0.057005,0.000124939,0.0177557,0.0177561,-0.0135363,0.0175524,7.93679e-05,0.00421425,0.00421499,-0.116581,0.0765708,-0.000297041,0.025051,0.0250528,-0.0404301,0.0564722,1.15489e-07,0.01239,0.01239,-0.0133247,0.0177102,7.48926e-05,0.0039964,0.0039971,-0.0138103,0.0130087,-9.61064e-05,0.00400004,0.0040012,-0.0125058,0.015983,6.50998e-05,0.00390336,0.00390391,-0.0417691,0.0633095,-4.81455e-05,0.0131107,0.0131107,1.03654,1.03655,1.37523,1.37538,1.34398,1.34405,1.0159,1.01596,1.00021,1.00022,1.03648,1.03648,1.35429,1.35432,28.689,28.689
421415583,3059364,1034193,0.00243048,137.746,-0.0140109,0.0180985,7.28386e-05,0.00406243,0.00406308,-0.217826,0.230758,0.000473817,0.067477,0.0674787,-0.692216,0.642741,-0.0017996,0.199148,0.199156,-0.0138542,0.0170776,6.82547e-05,0.00402982,0.0040304,-0.0150774,0.0125296,-0.000100538,0.00400434,0.0040056,-0.0138046,0.0178975,7.11706e-05,0.00402443,0.00402506,-0.692065,0.647335,-0.00176929,0.19962,0.199627,-0.0140438,0.0172841,6.44552e-05,0.00403549,0.00403601,-0.134259,0.151459,-0.000554154,0.0378288,0.0378329,-0.468386,0.385993,0.00124144,0.111354,0.111361,-0.0139263,0.0170207,6.73365e-05,0.00402912,0.00402969,-0.0150771,0.0125299,-0.000100518,0.00400436,0.00400562,-0.0138371,0.0170909,6.28681e-05,0.00399772,0.00399821,-0.468732,0.39245,0.0012037,0.111506,0.111513,1.00667,1.00671,1.78375,1.7836,1.78843,1.78839,1.00017,1.00018,0.999996,0.999996,1.00668,1.00671,1.79021,1.79017,29.067,29.067
6699655944,2428132460,465504975,0.048524,2.75918,-0.0211523,0.0338916,-7.19317e-05,0.00659268,0.00659307,-0.0523968,0.0529446,-0.000313313,0.0137879,0.0137915,-0.264123,0.251806,0.00200262,0.0678385,0.0678681,-0.0174284,0.015951,6.28474e-05,0.00436753,0.00436799,-0.0141906,0.0154551,5.92553e-05,0.00419188,0.0041923,-0.0150547,0.0232601,-3.23616e-05,0.0047731,0.00477321,-0.271027,0.265098,0.00227559,0.0711279,0.0711642,-0.0193924,0.0320042,-2.68767e-05,0.0062623,0.00626236,-0.0439764,0.0423448,-0.000194382,0.0126204,0.0126219,-0.237721,0.22182,0.00138901,0.0616495,0.0616652,-0.0177793,0.0158603,6.51105e-05,0.00436418,0.00436466,-0.0143085,0.0155233,6.03412e-05,0.004191,0.00419144,-0.0136536,0.0219361,-5.69043e-07,0.00454087,0.00454087,-0.243873,0.233025,0.00158645,0.064602,0.0646215,1.05276,1.05281,1.09251,1.09266,1.10039,1.10059,1.00077,1.00076,1.00021,1.00021,1.05114,1.05116,1.10102,1.10125,26.7503,26.7503
374818,56589,136,0.000315148,6.62351,-0.0141969,0.0155092,6.10016e-05,0.00432921,0.00432964,-0.050722,0.0483745,2.4543e-05,0.0148711,0.0148711,-1,19.9515,1.94686,3.81786,4.28559,-0.0141527,0.0151409,0.000767126,0.00387215,0.00394741,-0.00903232,0.045861,0.00469214,0.008305,0.00953882,-0.0126464,0.0179254,0.00066975,0.00430368,0.00435548,-1,20.158,1.95355,3.83113,4.30046,-0.014505,0.0145163,-1.16974e-05,0.0037325,0.00373252,-0.0208002,0.0125255,-0.000457887,0.00428552,0.00430991,-1,6.17705,0.174425,1.31119,1.32274,-0.0148625,0.0150774,5.15728e-05,0.00370825,0.00370861,-0.00984588,0.010098,-3.85997e-05,0.00293441,0.00293466,-0.0125355,0.0125707,-1.52336e-05,0.0032714,0.00327144,-1,6.17831,0.174925,1.31217,1.32378,1.15987,1.15998,3.47009,3.45045,2.91175,3.23993,1.0442,1.06439,2.83022,3.2504,1.31555,1.33137,2.9197,3.24863,36.889,36.889
540436824,847567,11153450,0.0201895,637.633,-0.0145555,0.017345,2.72715e-05,0.00410436,0.00410445,-0.488889,0.611286,-0.00216245,0.145983,0.145999,-0.0447843,0.0424222,2.80471e-05,0.0115995,0.0115995,-0.0144477,0.0166327,2.72871e-05,0.00402092,0.00402101,-0.0153895,0.0120004,-0.000126656,0.00400467,0.00400667,-0.0140172,0.0171134,2.39276e-05,0.00401941,0.00401948,-0.0469514,0.0428459,2.14111e-05,0.012369,0.012369,-0.0142108,0.0164645,3.51729e-05,0.00406742,0.00406757,-0.388008,0.440982,-0.00121166,0.116636,0.116642,-0.0316298,0.0310126,-4.41676e-05,0.0096015,0.0096016,-0.014018,0.0160679,3.35686e-05,0.00398981,0.00398995,-0.0153863,0.0119978,-0.000126621,0.00400457,0.00400657,-0.0136801,0.0162519,3.16581e-05,0.00398319,0.00398332,-0.0333747,0.0409735,-6.02701e-05,0.0103661,0.0103663,1.00908,1.00907,1.25162,1.25169,1.20809,1.20808,1.0078,1.00778,1.00002,1.00002,1.00909,1.00908,1.19322,1.1932,27.6107,27.6107
647883,2323,65699,0.0917706,278.899,-0.0134479,0.0142194,-0.000120936,0.00429006,0.00429176,-0.311096,0.369452,-0.000896059,0.098874,0.098878,-0.016817,0.0168884,4.25464e-05,0.00465209,0.00465228,-0.0115819,0.0131076,-0.000105884,0.00388093,0.00388238,-0.0103391,0.0137833,1.04924e-05,0.00297961,0.00297963,-0.0117899,0.0134808,-0.000108448,0.00388928,0.0038908,-0.021368,0.0200836,0.000166861,0.00618199,0.00618424,-0.0129949,0.0130809,-7.60657e-05,0.00413316,0.00413386,-0.246422,0.278426,0.00154649,0.07781,0.0778254,-0.0151368,0.0130023,-4.29074e-05,0.0041192,0.00411943,-0.0119715,0.0120286,-7.30129e-05,0.00376218,0.00376289,-0.0103626,0.0138402,1.13715e-05,0.00297972,0.00297974,-0.0120337,0.0120216,-6.77578e-05,0.00374802,0.00374863,-0.0179381,0.0190122,3.84218e-05,0.00548307,0.00548321,1.03796,1.0382,1.27071,1.27051,1.12937,1.12935,1.03156,1.03175,0.999962,0.999961,1.03769,1.03792,1.12747,1.12785,25.711,25.711
10933683,7343645,6343,0.000346922,1.48886,-0.0209748,0.0227613,-9.59846e-05,0.00617724,0.00617798,-0.0266317,0.0288881,-9.18179e-05,0.00806354,0.00806407,-1,26.8536,2.81279,5.11269,5.83536,-0.013297,0.0221683,0.00153492,0.00478782,0.00502785,-0.0129909,0.0280001,0.00233568,0.00541519,0.00589743,-0.015241,0.0231332,0.000881539,0.00503749,0.00511404,-1,27.1671,2.82564,5.13745,5.86324,-0.0132933,0.0144317,-9.16235e-06,0.0041656,0.00416561,-0.0148696,0.0138097,3.68052e-05,0.00420109,0.00420125,-1,9.42132,-0.04205,1.5133,1.51389,-0.0138653,0.0138436,-3.35375e-05,0.00407181,0.00407195,-0.0150874,0.0135626,4.84532e-07,0.00396661,0.00396661,-0.0105109,0.0099249,-5.28433e-06,0.00294898,0.00294898,-1,9.42626,-0.0412086,1.51518,1.51574,1.48292,1.48309,1.91939,1.91944,3.3785,3.85455,1.17585,1.23475,1.36519,1.48677,1.70822,1.73417,3.39065,3.86824,30.8583,30.8583
377724782,291621648,1717874,0.00255992,1.29526,-0.0224052,0.0205701,0.000135402,0.00634607,0.00634752,-0.0247779,0.0265473,1.37655e-05,0.00769737,0.00769738,-1,3.64383,0.105481,0.936648,0.942568,-0.0162028,0.0159496,0.000612338,0.00427182,0.00431548,-0.0144945,0.0205356,0.000631407,0.00447006,0.00451444,-0.0148413,0.0174127,0.000352219,0.00442033,0.00443434,-1,3.68136,0.107902,0.940734,0.946902,-0.0166347,0.0195343,0.00101125,0.00483311,0.00493777,-0.0193924,0.0177688,0.00114747,0.00538062,0.00550162,-1,1.50275,-0.198794,0.595615,0.627915,-0.0161995,0.0149185,0.000106656,0.00404212,0.00404353,-0.0180424,0.0120597,-2.34369e-05,0.00405026,0.00405033,-0.0110917,0.0103927,0.00055896,0.00325411,0.00330177,-1,1.51501,-0.198353,0.596529,0.628642,1.31304,1.2855,1.43057,1.39911,1.57257,1.50111,1.05683,1.06726,1.10365,1.11459,1.35838,1.34302,1.57701,1.50627,26.458,26.458
640041725,1131081,2138265,0.00332384,565.867,-0.0134259,0.0160926,-3.0529e-05,0.00411269,0.0041128,-0.463158,0.538495,-0.00237528,0.138157,0.138178,-0.286278,0.253617,0.00103029,0.0731798,0.0731871,-0.0135194,0.0162174,-2.69968e-05,0.00408837,0.00408846,-0.015741,0.0150982,-0.000147919,0.00410906,0.00411172,-0.0133543,0.0159974,-3.11256e-05,0.00409144,0.00409156,-0.285666,0.251532,0.00110006,0.0735907,0.0735989,-0.013408,0.0160636,-3.42303e-05,0.00409794,0.00409808,-0.368705,0.387296,-0.00465624,0.0966512,0.0967633,-0.197752,0.203655,0.00223688,0.0512678,0.0513166,-0.0133641,0.016091,-2.66682e-05,0.00408176,0.00408185,-0.0157412,0.0150996,-0.000147902,0.00410906,0.00411172,-0.0133365,0.0159685,-3.4808e-05,0.00407679,0.00407694,-0.193227,0.201695,0.0022958,0.0515786,0.0516297,1.0036,1.00359,1.42944,1.428,1.4274,1.42619,1.00162,1.00162,1,1,1.0036,1.00359,1.42677,1.42552,28.8097,28.8097
216843,206318,36525,0.0794564,1.05101,-0.0251475,0.0229097,0.000109256,0.00697607,0.00697693,-0.0282721,0.0244065,-4.3235e-05,0.00725462,0.00725475,-0.109901,0.123852,-3.32895e-05,0.0344724,0.0344724,-0.0121622,0.0128941,8.87071e-05,0.00352298,0.0035241,-0.011929,0.0127008,-4.17392e-05,0.00349995,0.0035002,-0.0151851,0.0119895,2.94883e-05,0.0037787,0.00377882,-0.117812,0.139311,4.11521e-05,0.0371832,0.0371832,-0.0157261,0.015996,0.000240621,0.00468747,0.00469364,-0.0185407,0.0179723,0.000104103,0.00485669,0.00485781,-0.0632455,0.0646761,-0.000806208,0.0182825,0.0183003,-0.0118909,0.0123085,8.97127e-05,0.00350868,0.00350983,-0.0121322,0.0128895,-3.28129e-05,0.00348856,0.00348871,-0.0107216,0.0103957,9.61715e-05,0.00280892,0.00281056,-0.0669512,0.0674166,-0.000878648,0.0193285,0.0193485,1.48824,1.48646,1.49374,1.49342,1.88554,1.88371,1.00408,1.00407,1.00327,1.00329,1.34525,1.34451,1.92375,1.92176,29.7007,29.7007
1237048,232484,778,0.00052914,5.321,-0.0165742,0.01871,-0.000129234,0.00477891,0.00478066,-0.0520426,0.0532504,3.8398e-05,0.0136916,0.0136916,-1,14.1745,1.12569,2.62859,2.85949,-0.0146716,0.0180697,0.000578369,0.00426577,0.0043048,-0.0117479,0.0497374,0.0037928,0.00756365,0.00846133,-0.0138034,0.0180725,0.00049299,0.00462061,0.00464683,-1,14.3559,1.13085,2.63805,2.87021,-0.0141355,0.0161781,-1.38577e-05,0.00410014,0.00410017,-0.0166761,0.0133292,0.000645777,0.00471398,0.004758,-1,3.93618,-0.173845,0.942148,0.958053,-0.0146716,0.0155402,-0.000123115,0.00405206,0.00405393,-0.0117789,0.0126591,6.37956e-05,0.00349918,0.00349976,-0.0121307,0.0138127,-1.53817e-06,0.00349933,0.00349933,-1,3.93901,-0.173352,0.943038,0.958839,1.16555,1.16597,2.90446,2.8776,2.79,2.98469,1.05274,1.06188,2.16155,2.41769,1.32042,1.32792,2.79739,2.99342,31.0963,31.0963
5886737105,227397359,44471331,0.00722101,25.8874,-0.0160813,0.0158528,-4.46173e-05,0.00454349,0.00454371,-0.121157,0.10091,6.01395e-05,0.0332021,0.0332022,-0.497736,0.624621,-0.000332927,0.168857,0.168858,-0.0165191,0.0150483,-4.6779e-05,0.0043191,0.00431936,-0.0167072,0.0129106,-4.15691e-06,0.00394645,0.00394646,-0.0153806,0.0154605,-4.28312e-05,0.00435067,0.00435088,-0.503887,0.622299,-6.13878e-05,0.170153,0.170153,-0.0158816,0.0161085,-4.94556e-05,0.00445705,0.00445732,-0.0931763,0.104126,-2.07427e-05,0.0272754,0.0272754,-0.522457,0.481332,8.09584e-05,0.138051,0.138051,-0.0165205,0.0151664,-4.84777e-05,0.00431812,0.00431839,-0.0167064,0.012905,-4.10682e-06,0.00394645,0.00394645,-0.0151897,0.0157049,-4.74537e-05,0.00426875,0.00426901,-0.52111,0.477521,0.000273968,0.139056,0.139057,1.01939,1.01938,1.21729,1.21729,1.22315,1.22315,1.00023,1.00022,1,1,1.01919,1.01918,1.22362,1.22362,26.3483,26.3483
\end{filecontents*}
\csvreader[before reading=\scriptsize, tabular=rrrrrrrrrrrrrrrr, table head= 
\toprule
&
\multicolumn{3}{c}{true cardinalities} &
\multicolumn{4}{c}{\acs{RMSE} inclusion-exclusion} &
\multicolumn{4}{c}{\acs{RMSE} maximum likelihood} &
\multicolumn{4}{c}{\acs{RMSE} improvement factor}
\\
\cmidrule(l){2-4}
\cmidrule(l){5-8}
\cmidrule(l){9-12}
\cmidrule(l){13-16}
& 
$|\symSetA|$ & 
$|\symSetB|$ & 
$|\symSetX|$ & 
$|\symSetA|$ &
$|\symSetB|$ & 
$|\symSetX|$ & 
$|\symSetU|$ & 
$|\symSetA|$ &
$|\symSetB|$ & 
$|\symSetX|$ & 
$|\symSetU|$ & 
$|\symSetA|$ &
$|\symSetB|$ & 
$|\symSetX|$ & 
$|\symSetU|$
\\\midrule,
table foot=\bottomrule
]
{joint_cardinality_calculation.csv}{
trueCardA=\trueCardA,
trueCardB=\trueCardB,
trueCardX=\trueCardX,
trueJaccardIdx=\trueJaccardIdx,
trueRatio=\trueRatio,
avgNumFunctionEvaluations=\avgNumFunctionEvaluations,
improvementRmseA=\improvementRmseA,
improvementRmseB=\improvementRmseB,
improvementRmseX=\improvementRmseX,
improvementRmseABX=\improvementRmseABX,
inclExclRMSEA=\inclExclRMSEA,
maxLikeRMSEA=\maxLikeRMSEA,
inclExclRMSEB=\inclExclRMSEB,
maxLikeRMSEB=\maxLikeRMSEB,
inclExclRMSEX=\inclExclRMSEX,
maxLikeRMSEX=\maxLikeRMSEX,
inclExclRMSEABX=\inclExclRMSEABX,
maxLikeRMSEABX=\maxLikeRMSEABX}{
\thecsvrow & 
\num{\trueCardA} & 
\num{\trueCardB} & 
\num{\trueCardX} &
\numformat{\inclExclRMSEA} & 
\numformat{\inclExclRMSEB} & 
\numformat{\inclExclRMSEX} & 
\numformat{\inclExclRMSEABX} & 
\numformat{\maxLikeRMSEA} & 
\numformat{\maxLikeRMSEB} & 
\numformat{\maxLikeRMSEX} & 
\numformat{\maxLikeRMSEABX} & 
\numformattwo{\improvementRmseA} &
\numformattwo{\improvementRmseB} &
\numformattwo{\improvementRmseX} &
\numformattwo{\improvementRmseABX}
}
\end{table*}
To evaluate the estimation error of the new joint cardinality estimation approach we have randomly generated \num{3000} independent pairs of sketches for which the relative complement cardinalities $|\symSetA|$ and $|\symSetB|$ and the intersection cardinality $|\symSetX|$ are known. Each pair is constructed by randomly generating three \ac{HLL} sketches filled with $|\symSetA|$, $|\symSetB|$, and $|\symSetX|$ distinct elements, respectively. Then we merged the first with the third and the second with the third to get sketches for $\symSetS_1 = \symSetA\cup\symSetX$ and $\symSetS_2=\symSetB\cup\symSetX$, respectively.

\cref{tbl:joint_estimation_results} shows the results for \ac{HLL} sketches with parameters $\symPrecision=16$ and $\symRegRange=16$. For different cardinality configurations $|\symSetA|$, $|\symSetB|$, and $|\symSetX|$ we have compared the conventional approach using the inclusion-exclusion principle with the new joint \ac{ML} approach. Among the considered cases there are also cardinalities that are small compared to the number of registers in order to prove that the new approach also covers the small cardinality range where many registers are still in initial state. We have determined the relative \ac{RMSE} from the estimates for $|\symSetA| = |\symSetS_1\setminus\symSetS_2|$, $|\symSetB| = |\symSetS_2\setminus\symSetS_1|$, and $|\symSetX| = |\symSetS_1\cap\symSetS_2|$ for all \num{3000} generated examples. In addition, we investigated the estimation error of $\symPoissonRateEstimate_\symSetASuffix+\symPoissonRateEstimate_\symSetBSuffix
+
\symPoissonRateEstimate_\symSetXSuffix$ for the union cardinality $|\symSetU| = |\symSetS_1\cup\symSetS_2|$. We also calculated the improvement factor which we defined as the \ac{RMSE} ratio between both approaches. Since we only observed values greater than one, the new \ac{ML} estimation approach improves the precision for all investigated cases. For some cases the improvement factor is even clearly greater than two. Due to the square root scaling of the error, this means that we would need four times more registers to get the same error when using the conventional approach. As the results suggest the new method works well over the full cardinality range without the need of special handling of small or large cardinalities. Obviously, the joint estimation algorithm is also able to reduce the estimation error for unions by a significant amount.

A reason why the \ac{ML} method performs better than the inclusion-exclusion method is that the latter only uses a fraction of the available information given by the sufficient statistic \eqref{equ:sufficient_joint_statistic}, because the corresponding estimator can be expressed as a function of just the three vectors $(\boldsymbol\symCountVariate^{<}_{1} + \boldsymbol\symCountVariate^{=} + \boldsymbol\symCountVariate^{>}_{1})$, 
$(\boldsymbol\symCountVariate^{<}_{2} + \boldsymbol\symCountVariate^{=} + \boldsymbol\symCountVariate^{>}_{2})$, and 
$(\boldsymbol\symCountVariate^{>}_{1} + \boldsymbol\symCountVariate^{=} + \boldsymbol\symCountVariate^{>}_{2})$. In contrast, the \ac{ML} method uses all the information as it incorporates each individual value of the sufficient statistic.

\section{Outlook}
\label{sec:future_work}

As we have shown, the \ac{ML} method is able to improve the cardinality estimates for results of set operations between two \ac{HLL} sketches. Unfortunately, joint cardinality estimation is much more expensive than for a single sketch, because it requires maximization of a multi-dimensional function. Since we have found the improved estimator which is almost as precise as the \ac{ML} estimator for the single sketch case, we could imagine that there also exists a faster algorithm for joint cardinality estimation of two sketches. It is expected that such a new algorithm makes use of all the information given by the sufficient statistic \eqref{equ:sufficient_joint_statistic}.

\begin{figure}
\centering
\includegraphics[width=\columnwidth]{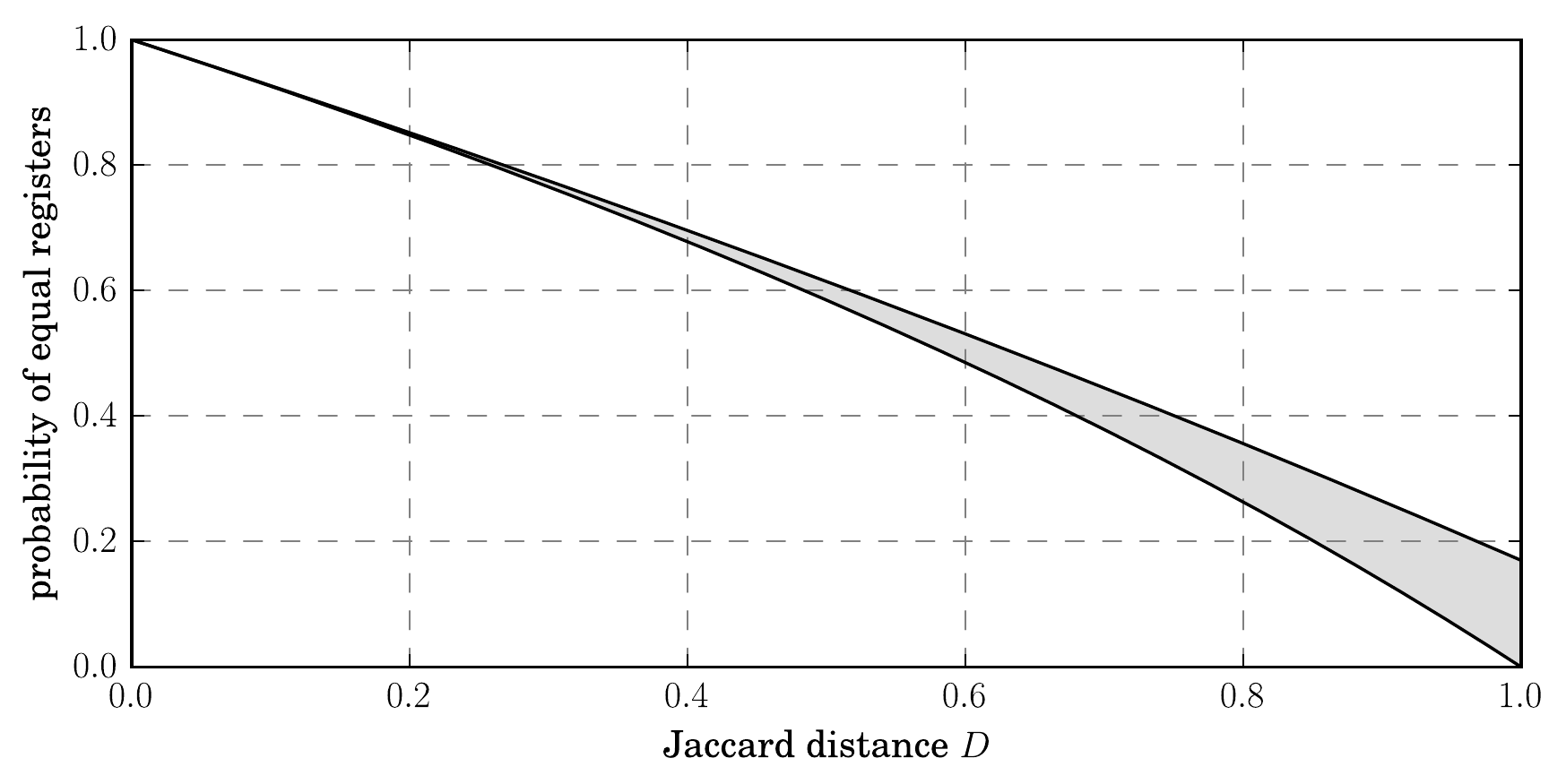}
\caption{The approximate probability range of equal register values as a function of the Jaccard distance.}
\label{fig:equal_register_probability}
\end{figure}

The \ac{ML} method can also be used to estimate distance measures such as the Jaccard distance of two sets that are represented as \ac{HLL} sketches. This directly leads to the question whether the \ac{HLL} algorithm could be used for locality-sensitive hashing \cite{Leskovec2014, Wang2014}. 
The \ac{HLL} algorithm itself can be regarded as hashing algorithm as it maps sets to register values. For sufficiently large cardinalities we can use the Poisson approximation and assume that the number of zero-valued \ac{HLL} registers can be ignored. Furthermore, if $\symPrecision+\symRegRange$ is chosen large enough, the number of saturated registers can be ignored as well. As a consequence, we can simplify \eqref{equ:joint_cdf} and assume that $\symRegValVariate_1$ and $\symRegValVariate_2$ are distributed according to 
\begin{equation*}
\symProbability(
\symRegValVariate_1 \leq \symRegVal_1
\wedge
\symRegValVariate_2 \leq \symRegVal_2
)
=
e^{
-
\frac{\symPoissonRate_{\symSetASuffix}}{\symNumReg 2^{\symRegVal_1}}
-
\frac{\symPoissonRate_{\symSetBSuffix}}{\symNumReg 2^{\symRegVal_2}}
-
\frac{\symPoissonRate_{\symSetXSuffix}}{\symNumReg 2^{\min(\symRegVal_1, \symRegVal_2)}}
}
\end{equation*}
which yields
\begin{multline*}
\symProbability(\symRegValVariate_1=\symRegValVariate_2)
=
\sum_{\symRegVal=-\infty}^\infty \symProbability(\symRegValVariate_1 = \symRegVal \wedge
\symRegValVariate_2 = \symRegVal)
=
\\
\sum_{\symRegVal = -\infty}^{\infty}
e^{-\frac{\symPoissonRate_\symSetASuffix + \symPoissonRate_\symSetBSuffix + \symPoissonRate_\symSetXSuffix}
{\symNumReg 2^\symRegVal}
}
\left(
1
-
e^{-\frac{\symPoissonRate_\symSetASuffix +  \symPoissonRate_\symSetXSuffix}
{\symNumReg 2^\symRegVal}
}
-
e^{-\frac{\symPoissonRate_\symSetBSuffix + \symPoissonRate_\symSetXSuffix}
{\symNumReg 2^\symRegVal}
}
+
e^{-\frac{\symPoissonRate_\symSetASuffix + \symPoissonRate_\symSetBSuffix + \symPoissonRate_\symSetXSuffix}
{\symNumReg 2^\symRegVal}
}
\right).
\end{multline*}
for the probability that a register has the same value in both sketches. Using the approximation 
$\sum_{\symRegVal=-\infty}^\infty
e^{-\frac{\symX}{2^\symRegVal}} - e^{-\frac{\symY}{2^\symRegVal}}
\approx
2\symAlpha_\infty (\log(\symY)-\log(\symX))$, which is obtained by integrating $\symPowerSeriesFunc(\symX)\approx 1$ on both sides, we get
\begin{equation}
\symProbability(\symRegValVariate_1=\symRegValVariate_2)
\approx
1
+
2\symAlpha_\infty
\log\left(
1
-
\textstyle\frac{1}{2}
\symDistanceMeasure
+
\textstyle\frac{1}{4}
\symDistanceMeasure^2
\textstyle\frac{
\symPoissonRate_\symSetASuffix
\symPoissonRate_\symSetBSuffix
}{\left(
\symPoissonRate_\symSetASuffix+
\symPoissonRate_\symSetBSuffix
\right)^2}
\right)
\end{equation}
where $\symDistanceMeasure = \frac{\symPoissonRate_\symSetASuffix
+\symPoissonRate_\symSetBSuffix}
{\symPoissonRate_\symSetASuffix+\symPoissonRate_\symSetBSuffix+
\symPoissonRate_\symSetXSuffix}$
 corresponds to the Jaccard distance. Since $
\frac{
\symPoissonRate_\symSetASuffix
\symPoissonRate_\symSetBSuffix
}{\left(
\symPoissonRate_\symSetASuffix+
\symPoissonRate_\symSetBSuffix
\right)^2}
$ is always in the range $[0,\frac{1}{4}]$, the probability for equal register values can be bounded by
$
1
+
2\symAlpha_\infty
\log(
1
-
\textstyle\frac{1}{2}
\symDistanceMeasure
)
\lesssim
\symProbability(\symRegValVariate_1=\symRegValVariate_2)
\lesssim
1
+
2\symAlpha_\infty
\log(
1
-
\textstyle\frac{1}{2}
\symDistanceMeasure
+
\textstyle\frac{1}{16}
\symDistanceMeasure^2
)
$ as shown in \cref{fig:equal_register_probability}.
This dependency on the Jaccard distance makes the
\ac{HLL} algorithm an interesting candidate for locality-sensitive hashing. 
Furthermore, the method described in \cref{sec:cardinality_estimation_set_intersections} would allow a more detailed estimation of the Jaccard distance by using the estimates for intersection and union sizes. This could be used for additional more precise filtering when searching for similar items.
Since \ac{HLL} sketches can be efficiently constructed, because only a single hash function evaluation is needed for each item, the preprocessing step would be very fast. In contrast, preprocessing is very costly for minwise hashing, because of the many required permutations \cite{Li2011}. 

\section{Conclusion}
\label{sec:conclusion}
Based on the Poisson approximation we have presented two new methods to estimate the number of distinct elements from \ac{HLL} sketches. Unlike previous approaches that 
use patchworks of different estimators or rely on empirically determined data, the presented ones are inherently unbiased over the full cardinality range. The first extends the original estimator by theoretically derived correction terms in order to obtain an improved estimator that can be straightforwardly translated into a fast algorithm. Due to its simplicity we believe that it has the potential to become the standard textbook cardinality estimator for \ac{HLL} sketches. The second estimation approach is based on the \ac{ML} method and can also be applied to two sketches which significantly improves cardinality estimates for corresponding intersections, relative complements, and unions compared to the conventional approach using the inclusion-exclusion principle.
\bibliographystyle{ACM-Reference-Format}
\bibliography{bibliography.bib} 

\end{document}